\documentclass{article}

 \usepackage[preprint]{neurips_2026}

\usepackage[utf8]{inputenc} % allow utf-8 input
\usepackage[T1]{fontenc}    % use 8-bit T1 fonts
\usepackage{hyperref}       % hyperlinks
\usepackage{url}            % simple URL typesetting
\usepackage{booktabs}       % professional-quality tables
\usepackage{amsfonts}       % blackboard math symbols
\usepackage{nicefrac}       % compact symbols for 1/2, etc.
\usepackage{microtype}      % microtypography
\usepackage{xcolor}         % colors
\usepackage{amsmath}
\usepackage{graphicx}
\usepackage{multirow}
\usepackage{enumitem}
\usepackage{wrapfig}
\usepackage{multicol}
\usepackage{amsthm}
\usepackage{tcolorbox}
\usepackage{wrapfig}
\tcbuselibrary{breakable}
\usepackage{subcaption}
\usepackage{float}
\usepackage{graphicx} 
\usepackage{algorithm}
\usepackage{algpseudocode}
\usepackage{caption}
\usepackage{graphicx}
\usepackage{amsmath,amssymb}  
\usepackage{amsthm} 
\usepackage{array}
\usepackage{makecell}
\usepackage{mathtools}

\setlist[itemize]{noitemsep,leftmargin=*,topsep=0pt}
\usepackage{xcolor}
\usepackage{ulem}

\title{Harness Engineering in LLM Tool Use via Agent-Native Reusable Tool Primitives}

\author{%
  Haibo Jin\\
  School of Information Sciences\\
  University of Illinois at Urbana-Champaign\\
  Champaign, IL 61820 \\
  \texttt{haibo@illinois.edu} \\ 
  \and
    \textbf{Suijin Wang} \\
 Independent Researcher\\
  Starc Institute \\
  \texttt{wangsuij22@gmail.com} \\
  \and
  \textbf{Xucheng Yu} \\
  School of Information Sciences\\
  University of Illinois at Urbana-Champaign\\
  Champaign, IL 61820 \\
  \texttt{xy63@illinois.edu} \\   \and
    \textbf{Haojing Luo} \\
 Independent Researcher\\
  Starc Institute \\
  \texttt{haojingluo9104@gmail.com} \\ 
  \and
  \textbf{Haohan Wang}\thanks{Corresponding Author} \\
  School of Information Sciences\\
  University of Illinois Urbana-Champaign\\
  Champaign, IL 61820 \\
  \texttt{haohanw@illinois.edu} \\}

\begin{document}

\maketitle

\begin{abstract}
    Large language models (LLMs) augmented with external tools have demonstrated remarkable capability in solving complex real-world tasks. However, existing approaches suffer from two key challenges: brittle multi-step and multi-turn reasoning caused by incompatible tool output types and API schemas, and performance degradation under large tool catalogues. To address these, we introduce \textbf{Tool Primitives}, a design that replaces rigid API schema-based invocation with natural language as the interface for tool calling, where each tool is wrapped with an LLM interface that handles schema resolution and execution internally, enabling natural inter-tool communication for nested and multi-turn tool calling. Building on Tool Primitives, we host \textbf{ToolFace}, a centralized repository of 25,519 functions from which LLMs dynamically retrieve only the relevant tools at inference time, eliminating the need to enumerate raw API schemas in context. To orchestrate Tool Primitives and ToolFace reliably in complex settings, we further propose \textbf{HEART}, a \textbf{H}arness \textbf{E}ngineering framework via \textbf{A}gent-native, \textbf{R}eusable \textbf{T}ool Primitives, comprising a Planner, Router, and Verifier that jointly support dynamic tool invocation planning, multi-step execution, and feedback-driven recovery.
    
    Experiments on five benchmarks demonstrate that HEART outperforms SFT-based models by $10\%$ on average and surpasses GPT-5.4, Claude-4.6-Sonnet, and Gemini-3.1-Pro by $6\%$ on average while reducing API cost by up to $85\%$. On 50 real-world tasks, HEART achieves $84\%$ task completion, $3.8\times$ the average of three frontier commercial models ($22\%$).
    
\end{abstract}

\section{Introduction}
Large Language Models (LLMs) with external tools and APIs has significantly enhanced the capability to solve complex real-world tasks~\cite{huang2024planning, qin2023toolllm}. To equip LLMs with tool-use capabilities, existing approaches broadly fall into two paradigms: training-free methods that elicit tool invocation through prompting and in-context demonstrations~\cite{yao2022react, shen2023hugginggpt}, and fine-tuning-based methods that train models on curated tool-use trajectories via supervised fine-tuning (SFT) and preference optimization~\cite{lin2024hammer, liu2024toolace, prabhakar2025apigen}. The latter have driven substantial improvements in structured function calling, with models increasingly evaluated on standardized benchmarks such as BFCLv4~\cite{patil2025bfcl}.

Despite rapid progress, existing LLM tool-use methods remain brittle in complex multi-step and multi-turn reasoning. When tool invocations involve dependent steps, models often fail to correctly map outputs from one tool into the expected input schema of the next, since tools expose heterogeneous APIs and return diverse output formats. This issue becomes even more severe in multi-turn interactions, where relevant context progressively degrades as histories grow longer. Existing benchmarks confirm these limitations: on nested API call benchmarks such as NESTFUL~\cite{basu2025nestful}, even the strongest models achieve only 28\% full-sequence match accuracy, while on multi-turn benchmarks such as $\tau^2$-Bench~\cite{barres2025tau}, performance drops substantially as interaction depth increases. This raises a natural question: \textbf{instead of requiring models to speak the language of APIs, can tools be made to speak the language of models?} We answer affirmatively with \textbf{Tool Primitives}, which replace rigid schema-based invocation with natural language interfaces for tool calling. Rather than requiring models to memorize exact parameter types and API schemas, each tool is wrapped with an LLM interface that handles schema resolution and execution internally. Models therefore interact with tools through natural language, making inter-tool communication more uniform and enabling nested tool calling and multi-turn information acquisition without explicit schema knowledge.

Second, a large tool catalogue degrades performance. Tool invocation is typically conditioned on a large catalogue of candidate tools provided in the input context; as the number of available tools grows, model performance degrades substantially. Recent evaluations show accuracy drops of 7\%-85\% as tool catalogue size scales from 8K to 120K tokens~\cite{kate2025longfunceval, liu2024lost}, and this problem is further amplified in real-world settings such as ToolBench~\cite{qin2023toolllm}, which encompasses over 16,000 real-world APIs. Motivated by this, we argue that \textbf{models need not know the full tool catalogue upfront and tools should instead be retrieved on demand at inference time}. Inspired by Agent Skills~\cite{anthropic2025agentskills} and Agent Primitives~\cite{jin2026agent}, both of which package instructions, metadata, and resources into modular reusable capability blocks, we treat tools analogously and introduce \textbf{ToolFace}, a large-scale tool repository containing 25,519 functions each paired with a structured schema and an executable implementation, where each function is further wrapped as a Tool Primitive to enable natural language interaction. When calling tools, LLMs dynamically search and retrieve only the relevant tools from ToolFace at inference time, directly avoiding the performance degradation caused by large tool catalogues.

To make Tool Primitives and ToolFace practical for real-world deployment, we introduce \textbf{HEART}, a \textbf{H}arness \textbf{E}ngineering via \textbf{A}gent-native, \textbf{R}eusable \textbf{T}ool Primitive. We design a multi-agent system (MAS) comprising a \textbf{Planner} that decomposes user queries into an invocation plan (or requests clarification when context is insufficient) and a \textbf{Router} that performs parameter mapping to bind function arguments and hyperparameters according to user queries before dispatching calls to the corresponding Tool Primitives. Tool Primitives further execute the tools and return results. We also employ a \textbf{Verifier} that evaluates each execution result against four criteria: task completion, argument consistency, execution validity, and constraint satisfaction. When verification fails, feedback is returned to the Planner for re-planning, enabling iterative recovery instead of one-shot execution.

Extensive experiments on five benchmarks spanning large-scale 
real-world API invocation across single- and multi-tool scenarios (ToolBench~\cite{qin2023toolllm}), nested API call sequences (NESTFUL~\cite{basu2025nestful}), dual-control conversational agent evaluation ($\tau^{2}$-Bench~\cite{barres2025tau}), robustness and fine-grained tool-use evaluation (ACEBench~\cite{chen2025acebench}) 
and agentic web search and memory function calling (BFCLv4~\cite{patil2025bfcl}), demonstrate that HEART consistently outperforms SFT-based models by $10\%$ on average and surpasses three commercial models, including GPT-5.4, Claude-4.6-Sonnet, and Gemini-3.1-Pro, by $6\%$ on average while reducing token cost by up to $85\%$. Furthermore, we curate 50 real-world tasks on which HEART achieves $84\%$ task completion and demonstrates robustness against prompt injection attacks, reducing attack success rate to $0.0\%$. Our main contributions are as follows:

\begin{itemize}
    \item We design \textbf{Tool Primitives}, a natural language interface that encapsulates schema resolution and execution within each tool, enabling nested and multi-turn tool calling without requiring explicit API schema knowledge. We further introduce \textbf{ToolFace}, a repository of 25,519 functions wrapped as Tool Primitives, enabling dynamic tool retrieval at inference time.
    
    \item We introduce \textbf{HEART},  a harness engineering framework for LLM tool use realized through a multi-agent system comprising a five-stage pipeline—Planning, Routing, Tools, Execution, and Verification—to jointly address key limitations of existing LLM tool-use approaches. 

    \item Experiments on five benchmarks show HEART outperforms SFT-based models by $10\%$ and three commercial models by $6\%$ on average, while reducing token cost by up to $85\%$. On 50 real-world tasks, HEART achieves $84\%$ task completion, $3.8\times$ the average of three frontier commercial models ($22\%$).
    
\end{itemize}

\section{Related Work}

\textbf{Tool-Use LLMs.} Prior work falls into two complementary axes: tool coverage and execution complexity.
% \textbf{Execution Complexity}
Early approaches to tool invocation remain confined to simple, stateless execution. For example, Hammer~\cite{lin2024hammer} focuses on efficient on-device deployment but operates within a single-turn regime where tools are invoked once and immediately resolved. Moving beyond isolated calls, MAGNET~\cite{yin2025magnetmultiturntoolusedata} synthesizes multi-turn trajectories via graph translation, using local dependency graphs and node operations (Insert, Merge, Split) to generate training data covering nested function compositions, long-range cross-turn dependencies, and missing-parameter scenarios. ToolACE~\cite{liu2024toolace} and the xLAM series~\cite{liu2024apigen} move beyond isolated single-turn queries, curating multi-turn conversational datasets that capture extended back-and-forth interactions between users and agents. While these efforts progressively advance from single-turn invocation toward sustained multi-turn execution, they share a common limitation: they lack explicit, structured mechanisms for planning, routing, and verifying complex tool execution paths, leaving robust multi-step composition an open challenge.

\textbf{Benchmarks for Tool Use Evaluation}. Evaluating LLM tool use spans several dimensions. BFCLv4~\cite{patil2025bfcl} evaluates structured function calling across diverse APIs, while StableToolBench~\cite{guo2024stabletoolbench} improves ToolBench~\cite{qin2023toolllm} by replacing unstable live APIs with deterministic virtual servers. NexusRaven~\cite{srinivasan2023nexusraven} focuses on schema-compliant JSON generation, API-Bank~\cite{li2023api} studies tool-augmented dialogue scenarios, and API-BLEND~\cite{basu2024api} evaluates compositional planning across heterogeneous APIs. NESTFUL~\cite{basu2025nestful} further targets nested API call sequences, showing that even GPT-4o achieves only 28\% full-sequence match accuracy. For interactive and multi-turn settings, $\tau$-Bench~\cite{yao2024tau} and $\tau^2$-Bench~\cite{barres2025tau} evaluate agents under rule-constrained conversational environments, while HammerBench~\cite{wang2025hammerbench}, ACEBench~\cite{chen2025acebench}, and ToolSandbox~\cite{lu2025toolsandbox} extend evaluation to mobile applications, complex multi-tool workflows, and stateful execution settings. 
% Web-based benchmarks including WebArena~\cite{zhou2023webarena}, MiniWoB++~\cite{humphreys2022data}, WebShop~\cite{yao2022webshop}, Mind2Web~\cite{deng2023mind2web}, and VisualWebArena~\cite{koh2024visualwebarena} evaluate browser-based interaction tasks ranging from form filling to multimodal web understanding. 
While these benchmarks cover diverse tool-use scenarios, they primarily evaluate task completion under fixed tool environments.

\textbf{Key Differences.} Aforementioned methods typically expose tools directly to the model as raw schemas within the input context, whether through prompting-based invocation or fine-tuning on static tool-use trajectories. This tightly couples tool storage, representation, and invocation, leading to the challenges discussed earlier. HEART departs from this paradigm in three ways. First, HEART introduces Tool Primitives, which provide tools with natural language interfaces instead of exposing raw API schemas directly to the model. Second, HEART hosts ToolFace, enabling retrieval of only relevant Tool Primitives at inference time rather than loading large tool catalogues into context. Third, HEART formulates tool use as a multi-agent process with planning, routing, execution, and verification stages, enabling iterative reasoning, multi-turn interaction, and recovery from failures.

\section{Methodology}\label{Methodology}
\subsection{Overview}
\begin{figure*}[htbp]
    \centering
    \includegraphics[width=1\linewidth]{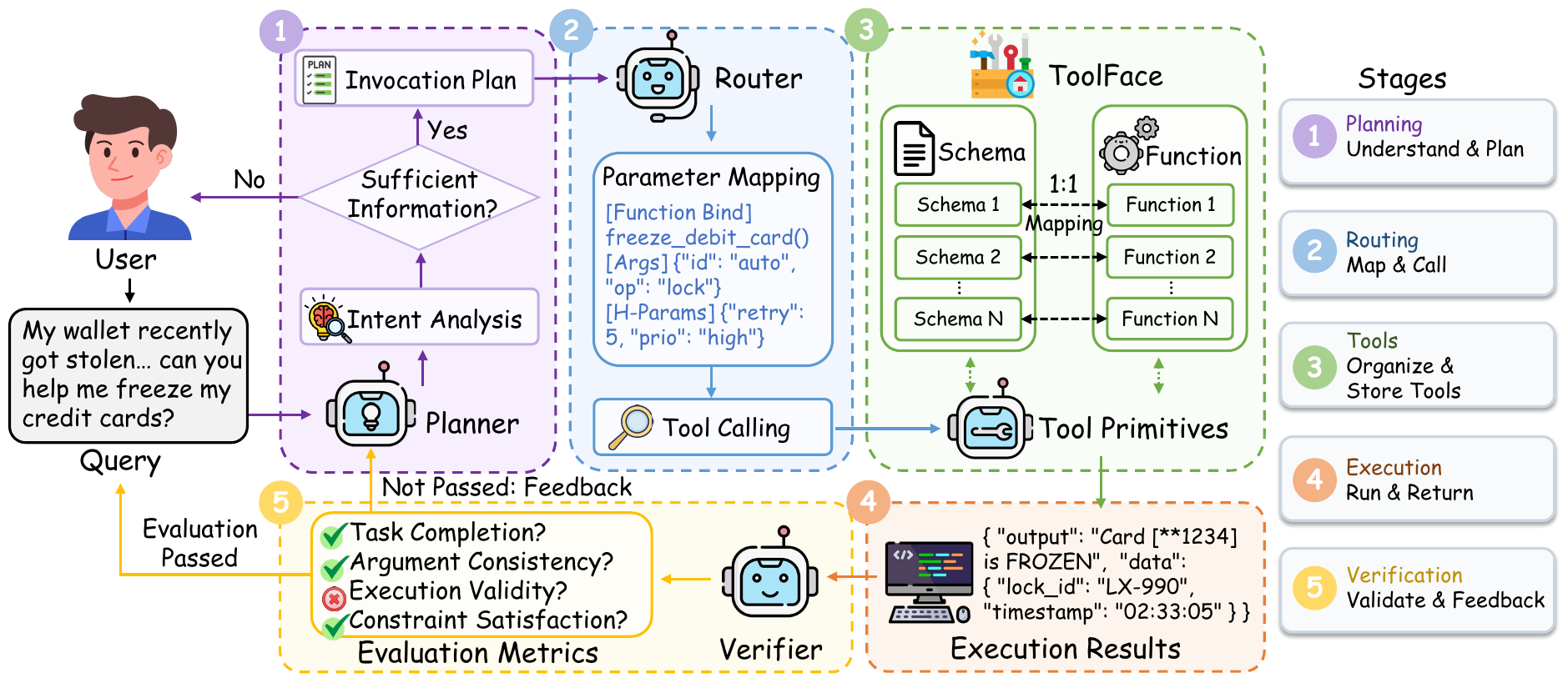}
    \vspace{-15pt}
    \caption{Overview of HEART. Planner performs intent analysis and checks for information sufficiency, soliciting missing details from the user if needed. Once sufficient context is gathered, the Router maps parameters and dispatches invocations to the corresponding Tool Primitives stored in ToolFace. Each Tool Primitive executes the underlying function and returns results to the Verifier, which evaluates task completion, argument consistency, execution validity, and constraint satisfaction. If verification fails, structured feedback is returned to the Planner for re-planning.}
    \label{pipline}
        \vspace{-10pt}
\end{figure*}

The core of HEART lies in two key innovations: \textbf{ToolFace}, a centralized tool repository that decouples tool storage from the model's input context, and \textbf{Tool Primitives}, LLM-wrapped interfaces that enable natural language invocation and inter-tool communication without requiring knowledge of raw API schemas. To make these two components work reliably in complex settings, we further introduce other supporting functions.
An overview of HEART is shown in Fig.~\ref{pipline}.

\subsection{ToolFace and Tool Primitives}

\textbf{Tool Primitives}. 
A Tool Primitive is an LLM-based wrapper that encapsulates a single tool from ToolFace, serving as its agent-native interface within the HEART pipeline. Rather than exposing raw API schemas directly to the calling model, each Tool Primitive accepts natural language invocation requests, internally resolves the corresponding schema from ToolFace, executes the underlying function, and returns a structured result. 

Formally, let $\mathcal{T} = \{(s_i, f_i)\}_{i=1}^{N}$ denote the ToolFace repository of $N$ schema--function pairs, where $s_i$ specifies the interface of tool $i$ and $f_i$ is its executable implementation. All Tool Primitives have a single base LLM 
$\mathcal{M}$, conditioned on different schemas at inference time. A Tool Primitive $\mathcal{P}_i$ is: 
\begin{equation}
    \mathcal{P}_i(x; c) = \mathcal{M}\bigl([s_i;\, c;\, x]\bigr),
\end{equation}
where $x$ is the natural language request issued by the upstream caller, $c$ is optional context (e.g., a prior primitive's result), $[\cdot]$ denotes prompt concatenation, and $c$ is left empty when no context is present. Upon receiving $x$ and $c$,
$\mathcal{P}_i$ interprets the request, maps it to the argument space defined by $s_i$, invokes $f_i$ with the resolved arguments, and returns the execution result $r_i$.

\textbf{ToolFace}. To decouple tools from the LLM's input and enable on-demand retrieval at inference time, we build ToolFace as a centralized tool registry. It stores tools as structured schema--function pairs: we manually author each schema to capture the tool's interface, including parameter types, constraints, and return specifications, and pair it with the corresponding executable implementation.

To build it, we collect 25,519 tools in total: 16,464 live APIs spanning 49 functional categories from ToolBench~\cite{qin2023toolllm}; 40 mathematical reasoning tools and 4,348 coding tools from NESTFUL~\cite{basu2025nestful}; 68 tools across retail, airline, and telecom domains from $\tau^{2}$-Bench~\cite{barres2025tau}; 4,538 bilingual APIs spanning 8 major domains and 68 sub-categories (including technology, finance, entertainment, and healthcare) from ACEBench~\cite{chen2025acebench}; 2 web search and 5 memory tools from BFCLv4~\cite{patil2025bfcl}; and 54 manually crafted tools to support human-like interactions such as form filling, button clicking, and webpage analysis.

% This design has three advantages. First, it enables \textbf{natural language invocation}: rather than requiring the Router to produce raw schema-compliant parameter dictionaries, each Tool Primitive accepts a natural language request $x_k$ and handles schema resolution and argument binding internally, decoupling the Router from low-level parameter formatting. Second, Tool Primitives support \textbf{inter-tool communication}: in sequential calling scenarios, the structured result $r_j$ of Primitive $\mathcal{P}_j$ is passed as natural language context directly into the invocation request of the next Primitive $\mathcal{P}_k$, enabling LLM-to-LLM dialogue across dependent steps without requiring the Router or Planner to manage intermediate state. Third, Tool Primitives provide \textbf{execution isolation}: each Primitive manages its own schema resolution and argument validation internally, so errors are localized and surfaced as structured failures to the Verifier rather than propagating through the pipeline.
% \hwc{this paragraph does not seem to belong in textual section. if you want to highlight these, it's probably better to put in results or discussions}

\subsection{Harness Engineering via Agent-Native Reusable Tool Primitives}

To make ToolFace and Tool Primitives work reliably in complex settings, we adopt harness engineering, realized through a multi-agent system with collaborative agents covering the full invocation lifecycle: a \textbf{Planner} for intent analysis and information sufficiency checking, a \textbf{Router} for parameter mapping and tool dispatch, and a \textbf{Verifier} for execution evaluation and feedback-driven recovery. We summarize their formal interfaces below; complete role specifications are provided in Appendix~\ref{appendix:agent-roles}, and prompts for each agent are listed in Appendix~\ref{prompt}.

\textbf{Planner}.
The Planner decomposes the user query $q$ into a structured intent and assesses whether the current context $\mathcal{C}_t$ is sufficient to construct an invocation plan:
\begin{equation}
    \delta_t = \textsc{Planner}(q, \mathcal{C}_t) \in \{\texttt{sufficient},\ \texttt{insufficient}\}.
\end{equation}
If $\delta_t = \texttt{insufficient}$, a targeted clarification request is returned to the user; otherwise, the Planner emits an invocation plan $\Pi = (\pi_1, \ldots, \pi_K)$, an ordered sequence of $K$ tool invocation steps. This iterative information acquisition mechanism directly addresses the multi-turn degradation challenge by resolving ambiguity before committing to execution, reducing downstream parameter mapping failures and invalid tool calls.

\textbf{Router}.
For each step $\pi_k \in \Pi$, the Router resolves required arguments from $\mathcal{C}_t$ and constructs a natural language invocation request together with execution-level configuration:
\begin{equation}
    \beta_k = \textsc{Router}(\pi_k,\ \mathcal{C}_t) = 
    \bigl(x_k,\ \texttt{H-Params}_k\bigr),
\end{equation}
where $x_k$ encodes the target tool, intended operation, and resolved parameter values, and $\texttt{H-Params}_k$ specifies execution hyperparameters such as retry count and priority level. The request $x_k$ is then dispatched to the corresponding Tool Primitive $\mathcal{P}_k$, which handles internal schema validation and argument binding. This indirection sidesteps two persistent failure modes of dispatching to raw tool implementations—fragile parameter formatting and degraded selection accuracy under large tool catalogues—directly motivating the Tool Primitive abstraction introduced earlier.

\textbf{Verifier}.
The Verifier evaluates each execution result $r_k$ along four criteria: \textbf{task completion}, \textbf{argument consistency}, \textbf{execution validity}, and \textbf{constraint satisfaction}, producing a verification:
\begin{equation}
    v_k = \textsc{Verifier}(r_k,\ \pi_k,\ s_k) \in \{\texttt{pass},\ \texttt{fail}\}.
\end{equation}
If $v_k = \texttt{pass}$, execution proceeds to $\pi_{k+1}$, or the final result is returned to the user when $k = K$. If $v_k = \texttt{fail}$, the Verifier generates structured feedback $\phi_k$ diagnosing the failure mode and returns it to the Planner to enrich the context and trigger targeted re-planning:
\begin{equation}
    \mathcal{C}_{t+1} = \mathcal{C}_t \cup \{\phi_k\},\quad 
    \Pi' = \textsc{Planner}(q,\ \mathcal{C}_{t+1}).
\end{equation}
This feedback-driven recovery loop directly addresses the troubleshooting challenge: rather than reverting to a brittle fallback upon tool failure, HEART systematically diagnoses the failure mode and re-plans with enriched context, enabling structured error recovery across the full invocation lifecycle. Detailed criterion definitions are provided in Appendix~\ref{appendix:verifier-criteria}. We also provide the end-to-end execution flow of HEART and the prompts for each agent in Appendix~\ref{end-to-end} and Appendix~\ref{prompt}, respectively.

\section{Experiments}\label{exp}
\subsection{Experimental Setup}\label{setup}

\textbf{Benchmarks.} We evaluate HEART on five benchmarks: ToolBench~\cite{qin2023toolllm} covers large-scale real-world API invocation across single- and multi-tool scenarios. NESTFUL~\cite{basu2025nestful} evaluates nested API call sequences where outputs of one call serve as inputs to the next. $\tau^2$-Bench~\cite{barres2025tau} assesses agents under dual-control conversational settings where both agent and user invoke tools in a shared environment. ACEBench~\cite{chen2025acebench} targets robustness and fine-grained tool-use evaluation across complex multi-tool workflows. BFCLv4~\cite{patil2025bfcl} focuses on agentic web search and memory function calling.

\textbf{Baselines.} Beyond benchmark-specific baselines, we include frontier models including GPT-5.4~\cite{openai2025gpt54}, Claude-4.6-Sonnet~\cite{anthropic2025claudesonnet46}, and Gemini-3.1-Pro~\cite{google2025gemini31} as shared baselines across all benchmarks.

\textbf{Implementation Details.} We use Qwen3-8B as the backbone LLM for all roles in HEART: Planner, Router, Verifier, and Tool Primitives, by default. All experiments are conducted with a maximum re-planning budget of $B = 3$. Tool retrieval from ToolFace is performed via semantic search over tool schema descriptors. We report average results over three runs.

\subsection{Main Results}

\textbf{On ToolBench.} 
We compare HEART against ToolBench~\cite{qin2023toolllm}'s baselines including Vicuna~\cite{zheng2023judging}, Alpaca~\cite{alpaca}, ToolLLaMA~\cite{qin2023toolllm}, GPT-3.5-Turbo~\cite{ouyang2022training}, Claude-2~\cite{claudesonnet2}, Text-Davinci-003~\cite{brown2020language}, and GPT-4~\cite{achiam2023gpt}. We report Pass Rate and Win Rate across six single- and multi-tool (I1, I2, I3) scenarios. 

% \wsj{Following ToolBench~\cite{qin2023toolllm}, we compare HEART against a wide spectrum of models (, ranging from Category ? models (Vicuna, Alpaca) and proprietary APIs). We report Pass Rate and Win Rate across six subsets that progress from single-tool (I1) to multi-tool (I2, I3) scenarios. }

HEART tops the ToolBench leaderboard (Table~\ref{toolbench_compare}) with an average Pass Rate of $75.1$\% and Win Rate of $75.7$\%. These figures represent gains of $+1.7$\%/$+1.9$\% over the strongest commercial baseline Claude-4.6-Sonnet (DFSDT) and $+7.8$\%/$+12.6$\% over the strongest fine-tuned baseline ToolLLaMA (DFSDT-Retriever).
The performance advantage of HEART is most pronounced on I3-Inst ($77.2$\% Pass, $89.4$\% Win), the hardest subset involving multi-tool instructions requiring cross-API dependencies, where HEART surpasses Claude-4.6-Sonnet with DFSDT by $+3.4$\% on Pass Rate and $+2.9$\% on Win Rate. This consistent advantage across difficulty levels demonstrates that HEART's structured Planner-Router-Verifier pipeline generalizes effectively to large-scale real-world API invocation without relying on explicit search strategies such as DFSDT.

\textbf{On NESTFUL.} Following NESTFUL~\cite{basu2025nestful}, we compare HEART against a broad set of baselines, including the xLAM series~\cite{zhang2024reverse, liu2024apigen}, Hammer~\cite{lin2024hammer}, ToolACE~\cite{liu2024toolace}, Llama-3.1~\cite{grattafiori2024llama}, Mixtral~\cite{jiang2024mixtral}, Granite-20B~\cite{abdelaziz2024granite}, DeepSeek-V3~\cite{liu2024deepseek}, and GPT-4o~\cite{hurst2024gpt}. We report F1 score for function name prediction (F1 Func.) and parameter name prediction (F1 Param.), partial sequence match accuracy (Part. Acc.), full sequence match accuracy (Full Acc.), and win rate (Win Rate), which 
measures whether all predicted APIs are executable and lead to the gold answer, under one-shot and three-shot ICL settings.

\begin{table*}[htbp]
\centering
\caption{Performance comparison (\%) of different methods on ToolBench.}
\vspace{-8pt}
\label{toolbench_compare}
\resizebox{\textwidth}{!}{%
\begin{tabular}{llcc|cc|cc|cc|cc|cc|cc}
\toprule
\multirow{2}{*}{\textbf{Model}} & \multirow{2}{*}{\textbf{Method}} 
& \multicolumn{2}{c}{\textbf{I1-Inst.}} 
& \multicolumn{2}{c}{\textbf{I1-Tool}} 
& \multicolumn{2}{c}{\textbf{I1-Cat.}} 
& \multicolumn{2}{c}{\textbf{I2-Inst.}} 
& \multicolumn{2}{c}{\textbf{I2-Cat.}} 
& \multicolumn{2}{c}{\textbf{I3-Inst.}} 
& \multicolumn{2}{c}{\textbf{Avg.}} \\
\cmidrule(lr){3-4} \cmidrule(lr){5-6} \cmidrule(lr){7-8}
\cmidrule(lr){9-10} \cmidrule(lr){11-12} \cmidrule(lr){13-14} \cmidrule(lr){15-16}
 & & \textbf{Pass} & \textbf{Win} & \textbf{Pass} & \textbf{Win} & \textbf{Pass} & \textbf{Win} & \textbf{Pass} & \textbf{Win} & \textbf{Pass} & \textbf{Win} & \textbf{Pass} & \textbf{Win} & \textbf{Pass} & \textbf{Win} \\
\midrule

Vicuna & ReACT \& DFSDT
& 0.0 & 0.0 & 0.0 & 0.0 & 0.0 & 0.0 & 0.0 & 0.0 & 0.0 & 0.0 & 0.0 & 0.0 & 0.0 & 0.0 \\

Alpaca & ReACT \& DFSDT
& 0.0 & 0.0 & 0.0 & 0.0 & 0.0 & 0.0 & 0.0 & 0.0 & 0.0 & 0.0 & 0.0 & 0.0 & 0.0 & 0.0 \\

\multirow{3}{*}{ToolLLaMA}
& ReACT
& 25.0 & 45.0 & 29.0 & 42.0 & 33.0 & 47.5 & 30.5 & 50.8 & 31.5 & 41.8 & 25.0 & 55.0 & 29.0 & 47.0 \\
& DFSDT
& 57.0 & 55.0 & 61.0 & 55.3 & 62.0 & 54.5 & 77.0 & 68.5 & 77.0 & 58.0 & 66.0 & 69.0 & 66.7 & 60.0 \\
& DFSDT-Retriever
& 64.0 & 62.3 & 64.0 & 59.0 & 60.5 & 55.0 & 81.5 & 68.5 & 68.5 & 60.8 & 65.0 & 73.0 & 67.3 & 63.1 \\

\midrule

\multirow{2}{*}{GPT-3.5-Turbo}
& ReACT
& 41.5 & -- & 44.0 & -- & 44.5 & -- & 42.5 & -- & 46.5 & -- & 22.0 & -- & 40.2 & -- \\
& DFSDT
& 54.5 & 60.5 & 65.0 & 62.0 & 60.5 & 57.3 & 75.0 & 72.0 & 71.5 & 64.8 & 62.0 & 69.0 & 64.8 & 64.3 \\

\multirow{2}{*}{Claude-2}
& ReACT
& 5.5 & 31.0 & 3.5 & 27.8 & 5.5 & 33.8 & 6.0 & 35.0 & 6.0 & 31.5 & 14.0 & 47.5 & 6.8 & 34.4 \\
& DFSDT
& 20.5 & 38.0 & 31.0 & 44.3 & 18.5 & 43.3 & 17.0 & 36.8 & 20.5 & 33.5 & 28.0 & 65.0 & 22.6 & 43.5 \\

\multirow{2}{*}{Text-Davinci-003}
& ReACT
& 12.0 & 28.5 & 20.0 & 35.3 & 20.0 & 31.0 & 8.5 & 29.8 & 14.5 & 29.8 & 24.0 & 45.0 & 16.5 & 33.2 \\
& DFSDT
& 43.5 & 40.3 & 44.0 & 43.8 & 46.0 & 46.8 & 37.0 & 40.5 & 42.0 & 43.3 & 46.0 & 63.0 & 43.1 & 46.3 \\

\multirow{2}{*}{GPT-4}
& ReACT
& 53.5 & 60.0 & 50.0 & 58.8 & 53.5 & 63.5 & 67.0 & 65.8 & 72.0 & 60.3 & 47.0 & 78.0 & 57.2 & 64.4 \\
& DFSDT
& 60.0 & 67.5 & 71.5 & 67.8 & 67.0 & 66.5 & 79.5 & 73.3 & 77.5 & 63.3 & 71.0 & 84.0 & 71.1 & 70.4 \\

\multirow{2}{*}{GPT-5.4}
& ReACT
& 54.5 & 62.0 & 52.8 & 58.5 & 54.2 & 63.5 & 65.4 & 65.8 & 71.5 & 62.0 & 46.5 & 76.5 & 57.5 & 64.7 \\
& DFSDT
& 61.2 & 67.5 & 64.5 & 65.2 & 63.8 & 64.2 & 76.5 & 71.2 & 76.2 & 65.8 & 68.4 & 81.2 & 68.8 & 69.2 \\

\multirow{2}{*}{Claude-4.6-Sonnet}
& ReACT
& 57.4 & 65.2 & 55.2 & 61.5 & 57.8 & 67.5 & 70.2 & 68.4 & 75.4 & 64.5 & 51.2 & 81.5 & 61.2 & 68.1 \\
& DFSDT
& 65.4 & 70.8 & 70.5 & 69.5 & 68.4 & 70.2 & 81.5 & 75.6 & 80.5 & 69.4 & 73.8 & 86.5 & 73.4 & 73.8 \\

\multirow{2}{*}{Gemini-3.1-Pro}
& ReACT
& 56.2 & 63.8 & 53.8 & 60.5 & 56.5 & 65.4 & 68.4 & 67.5 & 73.2 & 63.2 & 49.5 & 79.5 & 59.6 & 66.6 \\
& DFSDT
& 63.8 & 69.2 & 67.5 & 66.8 & 66.2 & 67.5 & 79.5 & 73.4 & 78.4 & 67.8 & 71.2 & 84.5 & 71.1 & 71.5 \\

\midrule
\textbf{HEART} & /
& \textbf{67.5} & \textbf{72.8}
& \textbf{71.5} & \textbf{70.8}
& \textbf{69.5} & \textbf{72.1}
& \textbf{83.4} & \textbf{77.5}
& \textbf{81.6} & \textbf{71.5}
& \textbf{77.2} & \textbf{89.4}
& \textbf{75.1} & \textbf{75.7} \\

\bottomrule
\end{tabular}
}
\vspace{-10pt}
\end{table*}

\begin{table*}[htbp]
\centering
% \vspace{-8pt}
\caption{Results on NESTFUL under one-shot and three-shot in-context learning (ICL).}
\vspace{-5pt}
\Huge
\label{NESTEDFUL_compare}
\resizebox{\textwidth}{!}{%
\begin{tabular}{llccccc|ccccc}
\toprule
\multirow{2}{*}{\textbf{Model}} & \multirow{2}{*}{\textbf{\#Params}} 
& \multicolumn{5}{c}{\textbf{One-shot ICL}} 
& \multicolumn{5}{c}{\textbf{Three-shot ICL}} \\
\cmidrule(lr){3-7} \cmidrule(lr){8-12}
& 
& \textbf{F1 Func.} & \textbf{F1-Param.} & \textbf{Part Acc.} & \textbf{Full Acc.} & \textbf{Win Rate}
& \textbf{F1 Func.} & \textbf{F1-Param.} & \textbf{Part Acc.} & \textbf{Full Acc.} & \textbf{Win Rate}\\
\midrule

xLAM-1b-fc-r & 1B & 0.19 & 0.08 & 0.09 & 0.00 & 0.01 & 0.22 & 0.09 & 0.09 & 0.03 & 0.02 \\
xLAM-2-1b-fc-r & 1B & 0.41 & 0.13 & 0.13 & 0.00 & 0.00 & 0.43 & 0.12 & 0.13 & 0.00 & 0.00 \\
xLAM-7b-fc-r & 7B & 0.49 & 0.17 & 0.15 & 0.00 & 0.03 & 0.55 & 0.23 & 0.23 & 0.15 & 0.14 \\
xLAM-2-8b-fc-r & 8B & 0.48 & 0.15 & 0.14 & 0.00 & 0.01 & 0.47 & 0.17 & 0.15 & 0.04 & 0.04 \\

Hammer2.0-7b & 7B & 0.56 & 0.24 & 0.21 & 0.07 & 0.16 & 0.61 & 0.30 & 0.29 & 0.22 & 0.25 \\
Hammer2.1-7b & 7B & 0.10 & 0.05 & 0.05 & 0.01 & 0.01 & 0.16 & 0.10 & 0.11 & 0.08 & 0.08 \\

Llama-3.1-8B-Instruct & 8B & 0.64 & 0.19 & 0.17 & 0.06 & 0.06 & 0.63 & 0.22 & 0.22 & 0.16 & 0.11 \\
ToolACE-8B & 8B & 0.43 & 0.13 & 0.13 & 0.00 & 0.00 & 0.50 & 0.15 & 0.13 & 0.00 & 0.00 \\
ToolACE-2-Llama-3.1-8B & 8B & 0.28 & 0.10 & 0.13 & 0.00 & 0.00 & 0.29 & 0.10 & 0.13 & 0.00 & 0.00 \\

Granite-20B-FunctionCalling & 20B & 0.64 & 0.20 & 0.17 & 0.02 & 0.05 & 0.61 & 0.25 & 0.26 & 0.21 & 0.20 \\
Mixtral-8x7B-Instruct-v0.1 & 46.7B & 0.22 & 0.07 & 0.05 & 0.00 & 0.01 & 0.32 & 0.13 & 0.14 & 0.09 & 0.09 \\
xLAM-8x7b-fc-r & 46.7B & 0.40 & 0.15 & 0.16 & 0.01 & 0.01 & 0.43 & 0.16 & 0.17 & 0.02 & 0.03 \\

Llama-3.1-70B-Instruct & 70B & 0.41 & 0.19 & 0.15 & 0.04 & 0.09 & 0.33 & 0.17 & 0.15 & 0.07 & 0.11 \\
Mixtral-8x22B-Instruct-v0.1 & 141B & 0.49 & 0.21 & 0.17 & 0.06 & 0.07 & 0.65 & 0.29 & 0.28 & 0.21 & 0.23 \\
xLAM-8x22b-fc-r & 141B & 0.53 & 0.21 & 0.22 & 0.12 & 0.03 & 0.50 & 0.23 & 0.25 & 0.17 & 0.06 \\

Llama-3.1-405B-Instruct-fp8 & 405B & 0.41 & 0.14 & 0.08 & 0.03 & 0.10 & 0.41 & 0.18 & 0.13 & 0.07 & 0.14 \\
DeepSeek-V3 & 685B & 0.69 & 0.36 & 0.27 & 0.09 & 0.43 & 0.69 & 0.42 & 0.37 & 0.29 & 0.60 \\

\midrule
GPT-4o (2024-08-06) & UNK & 0.73 & 0.41 & 0.38 & 0.28 & 0.59 & 0.74 & 0.41 & 0.38 & 0.28 & 0.60 \\
GPT-5.4 & UNK & 0.78 & 0.48 & 0.50 & 0.40 & 0.66 & 0.80 & 0.50 & 0.52 & 0.42 & 0.68 \\
Claude-4.6-Sonnet & UNK & 0.77 & 0.47 & 0.49 & 0.39 & 0.65 & 0.79 & 0.49 & 0.51 & 0.41 & 0.67 \\
Gemini-3.1-Pro & UNK & 0.76 & 0.46 & 0.48 & 0.38 & 0.63 & 0.78 & 0.48 & 0.50 & 0.40 & 0.65 \\

\midrule
\textbf{HEART} & 8B * 4
& \textbf{0.82} & \textbf{0.52} & \textbf{0.55} & \textbf{0.44} & \textbf{0.75} 
& \textbf{0.84} & \textbf{0.56} & \textbf{0.58} & \textbf{0.47} & \textbf{0.79} \\

\bottomrule
\end{tabular}
}
\vspace{-12pt}
\end{table*}

Table~\ref{NESTEDFUL_compare} shows that HEART achieves the best performance on all five metrics under both one-shot and three-shot ICL.
Against the strongest commercial baseline GPT-5.4, HEART improves Full Acc. by $+4$\%/$+5$\% and Win Rate by $+9$\%/$+11$\% under one-shot/three-shot.
Strikingly, several tool-calling models, including all xLAM variants up to 8B and both ToolACE models, achieve Full Acc. $=0.00$ under one-shot, confirming that flat tool-use trajectories do not generalize to nested API sequences.
We also observe that raw model scale does not compensate for the absence of structured planning: DeepSeek-V3 (685B) reaches only $0.09$ Full Acc., far below HEART's $0.44$.

\textbf{On $\tau^2$-Bench.} Following the $\tau^2$-Bench~\cite{barres2025tau}, we include GPT-4.1~\cite{openai2025gpt41}, GPT-o4-mini~\cite{openai2025gpto4}, GPT-4.1-mini~\cite{openai2025gpt41}, Claude-3.7-Sonnet~\cite{anthropic2025claude37sonnet} and report Pass$^k$ ($k = 1, 2, 3, 4$), which measures the fraction of tasks completed successfully within $k$ interaction rounds, alongside average per task token consumption, API cost (USD) (Detailed in Appendix~\ref{appendix:pricing}), and latency (seconds). 

\begin{table*}[htbp]
\centering
\caption{Performance comparison of different methods across domains on the $\tau^2$-Bench benchmark.}
\vspace{-6pt}
\label{t2_compare}
\resizebox{0.85\linewidth}{!}{%
\begin{tabular}{llcccc|cc}
\toprule
\textbf{Method} & \textbf{Domain} & \textbf{Pass$^1$} & \textbf{Pass$^2$} & \textbf{Pass$^3$} & \textbf{Pass$^4$} & \textbf{Tokens / Cost (\$)} & \textbf{Latency (s)} \\
\midrule

\multirow{3}{*}{GPT-4.1}
& Retail  & 0.75 & 0.64 & 0.58 & 0.53 & 7,401 / 0.0592 & 14.28 \\
& Airline & 0.56 & 0.45 & 0.42 & 0.40 & 7,825 / 0.0626 & 15.62 \\
& Telecom & 0.34 & 0.26 & 0.22 & 0.19 & 6,862 / 0.0549 & 34.41 \\

\midrule
\multirow{3}{*}{GPT-o4-mini}
& Retail  & 0.71 & 0.59 & 0.52 & 0.46 & 5127 / 0.0031 & 4.24 \\
& Airline & 0.59 & 0.48 & 0.42 & 0.38 & 5389 / 0.0032 & 4.68 \\
& Telecom & 0.42 & 0.33 & 0.29 & 0.26 & 4856 / 0.0029 & 26.05 \\

\midrule
\multirow{3}{*}{GPT-4.1-mini}
& Retail  & 0.66 & 0.53 & 0.44 & 0.39 & 5,214 / 0.0083 & 6.12 \\
& Airline & 0.51 & 0.39 & 0.32 & 0.26 & 5,563 / 0.0089 & 6.81 \\
& Telecom & 0.44 & 0.30 & 0.22 & 0.18 & 5,021 / 0.0080 & 21.94 \\

\midrule
\multirow{3}{*}{Claude-3.7-sonnet}
& Retail  & 0.79 & 0.69 & 0.69 & 0.60 & 6,482 / 0.0972 & 11.45 \\
& Airline & 0.50 & 0.41 & 0.38 & 0.36 & 6,745 / 0.1012 & 12.32 \\
& Telecom & 0.49 & 0.37 & 0.31 & 0.25 & 6,038 / 0.0906 & 30.88 \\

\midrule
\multirow{3}{*}{GPT-5.4}
& Retail  & 0.70 & 0.53 & 0.43 & 0.37 & 7,842 / 0.1176 & 16.24 \\
& Airline & 0.64 & 0.55 & 0.51 & 0.48 & 8,126 / 0.1219 & 18.42 \\
& Telecom & 0.56 & 0.40 & 0.34 & 0.29 & 7,319 / 0.1098 & 35.11 \\

\midrule
\multirow{3}{*}{Claude-4.6-sonnet}
& Retail  & 0.79 & 0.73 & 0.65 & 0.60 & 7,153 / 0.1073 & 14.12 \\
& Airline & 0.71 & 0.64 & 0.60 & 0.51 & 7,412 / 0.1112 & 16.52 \\
& Telecom & 0.58 & 0.49 & 0.37 & 0.33 & 6,728 / 0.1009 & 33.48 \\

\midrule
\multirow{3}{*}{Gemini-3.1-Pro}
& Retail  & 0.68 & 0.54 & 0.46 & 0.40 & 6,914 / 0.0830 & 9.15 \\
& Airline & 0.54 & 0.41 & 0.33 & 0.28 & 7,258 / 0.0871 & 10.82 \\
& Telecom & 0.55 & 0.43 & 0.36 & 0.31 & 6,492 / 0.0779 & 28.46 \\

\midrule
\multirow{3}{*}{\textbf{HEART}}
& Retail  & \textbf{0.82} & \textbf{0.79} & \textbf{0.77} & \textbf{0.73} & 21,684 / 0.0152 & 18.72 \\
& Airline & \textbf{0.72} & \textbf{0.69} & \textbf{0.65} & \textbf{0.63} & 24,531 / 0.0172 & 21.34 \\
& Telecom & \textbf{0.62} & \textbf{0.59} & \textbf{0.55} & \textbf{0.50} & 20,887 / 0.0146 & 37.95 \\

\bottomrule
\end{tabular}}
\vspace{-8pt}
\end{table*}

% Table~\ref{t2_compare} reports results across three domains: Retail, Airline, and Telecom. HEART achieves the best Pass$^k$ scores across all domains and all $k$, outperforming the strongest commercial baseline, Claude-4.6-Sonnet, by up to $+51.5\%$ on Pass$^4$ in the Telecom domain, and by $+29.2\%$ on average Pass$^4$ across all three domains. Notably, the performance gap widens as $k$ increases: while Pass$^1$ margins over Claude-4.6-Sonnet remain relatively modest ($+1.4\%$--$+6.9\%$), HEART maintains substantially higher Pass$^4$ scores, demonstrating that its iterative re-planning and feedback-driven recovery sustain task completion under extended multi-turn interactions where commercial models degrade. Telecom consistently poses the greatest challenge for all methods, with the best commercial baseline reaching only $0.33$ on Pass$^4$, while HEART achieves $0.50$, reflecting HEART's stronger handling of complex tool dependencies and longer interaction horizons. Despite consuming more tokens due to its multi-agent architecture, HEART reduces API cost by up to $7.4\times$ relative to GPT-5.4 and $6.8\times$ relative to Claude-4.6-Sonnet, demonstrating that strong multi-turn tool-use performance need not come at high inference cost.

Table~\ref{t2_compare} shows that HEART achieves the best Pass$^k$ scores across all domains and all $k$.
The performance gap widens as $k$ increases: while Pass$^1$ margins over the strongest commercial baseline Claude-4.6-Sonnet remain modest, HEART maintains higher Pass$^4$ scores (e.g., $+51.5$\% in Telecom), demonstrating that iterative re-planning sustains task completion under extended multi-turn interactions where commercial models degrade.
Telecom consistently poses the greatest challenge; the best baseline reaches only $0.33$ on Pass$^4$, while HEART achieves $0.50$.
Despite higher token consumption from its multi-agent architecture, HEART reduces API cost by up to $7.4\times$ relative to GPT-5.4.

\textbf{On ACEBench.} We report performance across six Normal subcategories (Atom, Single-Turn, Multi-Turn, Similar API, Preference, Summary), the Special category, the Agent category, and an Overall score. We compare HEART against commercial models in Table~\ref{ACEBench_compare}; results against open-source baselines including Qwen2.5 series~\cite{qwen2.5}, Llama-3.1~\cite{grattafiori2024llama}, DeepSeek-V3~\cite{liu2024deepseek}, Phi-3~\cite{abdin2024phi3technicalreporthighly}, and tool-calling models such as Hammer~\cite{lin2024hammer}, xLAM~\cite{liu2024apigen}, and Watt-Tool-8B~\cite{wattai2024watttool8b} are in Appendix~\ref{appendix:acebench-open}.

\begin{table*}[htbp]
\centering
\vspace{-8pt}
\caption{Performance comparison (\%) of different methods across domains on the ACEBench.}
\vspace{-6pt}
\Huge
\label{ACEBench_compare}
\resizebox{\linewidth}{!}{%
\begin{tabular}{lcccccc|ccc}
\toprule
\multirow{2}{*}{\textbf{Model}} 
& \multicolumn{6}{c}{\textbf{Normal}} 
& \multirow{2}{*}{\textbf{Special}} 
& \multirow{2}{*}{\textbf{Agent}} 
& \multirow{2}{*}{\textbf{Overall}} \\
\cmidrule(lr){2-7}
& \textbf{Atom} & \textbf{Single-Turn} & \textbf{Multi-Turn} & \textbf{Similar API} & \textbf{Preference} & \textbf{Summary} & & & \\

\midrule
GPT-4o                          & 93.4 & 84.5 & 77.0 & 85.0 & 83.0 & 87.6 & 93.0 & 63.8 & 85.4 \\
GPT-4-Turbo                     & 93.2 & 84.8 & 77.5 & 86.0 & 86.0 & 88.0 & 86.7 & 67.5 & 84.5 \\
Qwen-Max                        & 91.2 & 80.5 & 68.0 & 83.0 & 83.0 & 84.2 & 74.0 & 64.3 & 78.4 \\
GPT-4o-Mini                     & 86.5 & 76.0 & 66.5 & 77.0 & 78.0 & 79.9 & 79.0 & 33.3 & 72.5 \\
Gemini-1.5-Pro                  & 84.5 & 76.8 & 64.5 & 80.0 & 78.0 & 79.0 & 78.7 & 25.5 & 70.7 \\
Claude-3.5-Sonnet               & 76.9 & 72.5 & 62.5 & 71.0 & 72.0 & 72.9 & 77.4 & 39.5 & 68.9 \\
GPT-5.4                         & 94.0 & 86.0 & 81.0 & 88.0 & 85.0 & 89.5 & 94.0 & 69.5 & 86.0 \\
Claude-4.6-Sonnet               & 93.8 & 85.5 & 79.5 & 87.0 & 84.0 & 88.3 & 91.3 & 66.8 & 84.3 \\
Gemini-3.1-Pro                      & 92.5 & 83.0 & 76.0 & 85.0 & 82.0 & 86.6 & 89.0 & 64.5 & 82.3 \\
\midrule
\textbf{HEART}                  & \textbf{94.2} & \textbf{87.5} & \textbf{83.5} & \textbf{89.0} & \textbf{86.0} & \textbf{89.8} & \textbf{95.0} & \textbf{72.3} & \textbf{86.9} \\
\bottomrule
\end{tabular}
}
\vspace{-8pt}
\end{table*}

Table~\ref{ACEBench_compare} shows that HEART achieves the highest Overall score ($86.9$\%), surpassing GPT-5.4 ($86.0$\%) by $+0.9$\%. While margins on Normal subcategories are modest, the clearest advantages appear on Special ($95.0$\% vs. $94.0$\%) and Agent ($72.3$\% vs. $69.5$\%), which require handling out-of-distribution tool formats and multi-step agentic reasoning. On Multi-Turn ($83.5$\%), HEART outperforms GPT-5.4 by $+2.5$\%, reflecting the benefit of iterative re-planning.

% Table~\ref{ACEBench_compare} reports results across all models. HEART achieves the highest Overall score of $86.9\%$, surpassing the strongest commercial baseline GPT-5.4 ($86.0\%$) by $+0.9$\%. While margins over commercial models are modest on Normal subcategories, HEART shows its clearest advantages on the two most challenging categories: Special ($95.0\%$ vs. $94.0\%$, $+1.0$\% over GPT-5.4) and Agent ($72.3\%$ vs. $69.5$\%, $+2.8$\% over GPT-5.4), which require handling out-of-distribution tool formats and multi-step agentic reasoning, respectively. The advantage is most pronounced on Multi-Turn ($83.5\%$), where HEART outperforms GPT-5.4 by $+2.5$\% and all open-source models by a large margin — $+12.5$\% over the best open-source baseline Qwen2.5-Coder-32B-Instruct ($71.0\%$) — reflecting the benefit of HEART's iterative re-planning mechanism in sustained multi-turn interactions. In contrast, SFT-based tool-calling models struggle severely on Special and Agent categories: Watt-Tool-8B achieves only $6.0\%$ on Special and $2.8\%$ on Agent, and xLAM-7B-r achieves $0.0$ on Preference, confirming that models fine-tuned on narrow tool-use trajectories fail to generalize to robustness-critical and agentic scenarios. Overall, HEART outperforms the best open-source model by $+7.3$\% on the Overall score, demonstrating that it can consistently improve robustness and generalization in complex tool-use settings.

\textbf{On BFCLv4.} 
Following BFCLv4~\cite{patil2025bfcl}, we compare HEART against a diverse set of baselines, including the xLAM series~\cite{liu2024apigen}, Hammer~\cite{lin2024hammer}, LLaMA models~\cite{grattafiori2024llama}, Qwen3~\cite{yang2025qwen3}, GPT-4.1~\cite{openai2025gpt41}, GPT-4.1-Mini~\cite{openai2025gpt41}, Gemini-1.5-Pro~\cite{team2024gemini}, and Claude-3.5-Sonnet~\cite{claudesonnet35}. We evaluate Web Search function calling under Base and No-Snippet settings, and Memory function calling across KV Retrieval, Vector Retrieval, and Recursive Summarization, reporting the averaged Overall score for each.

\begin{table*}[htbp]
\centering
\caption{Performance comparison (\%) of different methods on BFCLv4.}
\vspace{-6pt}
\label{BFCLv4_compare}
\resizebox{0.9\linewidth}{!}{%
\begin{tabular}{lccc|cccc}
\toprule
\multirow{2}{*}{\textbf{Model}} 
& \multicolumn{3}{c}{\textbf{Web Search}} 
& \multicolumn{4}{c}{\textbf{Memory}} \\
\cmidrule(lr){2-4} \cmidrule(lr){5-8}
& \textbf{Base} & \textbf{No Snippet} & \textbf{Overall} 
& \textbf{KV} & \textbf{Vector} & \textbf{Recursive Sum} & \textbf{Overall} \\
\midrule

xLAM-2-32B-FC-R       & 37.00 & 14.00 & 25.50 & 6.45  & 10.32 & 45.81 & 20.86 \\
xLAM-2-70B-FC-R       & 15.00 & 17.00 & 13.00 & 2.58  & 10.97 & 29.68 & 14.41 \\
xLAM-2-8B-FC-R        & 11.00 & 2.00  & 6.50  & 5.81  & 15.48 & 20.65 & 13.98 \\
xLAM-2-3B-FC-R        & 3.00  & 2.00  & 2.50  & 5.81  & 5.81  & 22.58 & 11.40 \\
xLAM-2-1B-FC-R        & 0.00  & 0.00  & 0.00  & 3.87  & 3.87  & 3.87  & 3.87  \\

Hammer2.1-7B          & 0.00  & 0.00  & 0.00  & 0.00  & 0.00  & 0.00  & 0.00  \\
Hammer2.1-3B          & 0.00  & 0.00  & 0.00  & 2.58  & 3.87  & 2.58  & 3.01  \\

LLaMA-3.1-8B-Instruct & 6.00  & 0.00  & 3.00  & 7.74  & 5.81  & 18.71 & 10.75 \\
LLaMA-3.2-3B-Instruct & 2.00  & 0.00  & 1.00  & 3.23  & 3.23  & 12.26 & 6.24  \\

Qwen3-8B              & 15.00 & 9.00  & 12.00 & 5.16  & 7.10  & 31.61 & 14.62 \\
Qwen3-32B             & 25.00 & 18.00 & 21.50 & 12.26 & 25.81 & 41.94 & 26.67 \\

\midrule

GPT-4.1               & 67.00 & 69.00 & 68.00 & 16.13 & 18.06 & 37.42 & 23.87 \\
GPT-4.1-Mini          & 62.00 & 52.00 & 57.00 & 22.58 & 16.13 & 41.94 & 26.88 \\
Gemini-1.5-Pro        & 60.00 & 64.00 & 62.00 & 19.35 & 24.52 & 58.71 & 34.19 \\
Claude-3.5-Sonnet     & 60.00 & 64.00 & 62.00 & 13.55 & 47.10 & 55.48 & 38.71 \\

GPT-5.4               & 81.00 & 83.00 & 82.00 & 57.42 & 58.71 & 51.61 & 55.91 \\
Claude-4.6-Sonnet     & 85.00 & 81.00 & 83.00 & 64.19 & 67.42 & 73.23 & 68.28 \\
Gemini-3.1-Pro            & 80.00 & 78.00 & 82.00 & 59.35 & 62.58 & 63.23 & 61.72 \\

\midrule
\textbf{HEART}        & \textbf{87.00} & \textbf{85.00} & \textbf{86.00} 
                      & \textbf{68.47} & \textbf{69.10} & \textbf{70.58} & \textbf{69.38} \\

\bottomrule
\end{tabular}}
\vspace{-5pt}
\end{table*}

Table~\ref{BFCLv4_compare} reports results across all models. HEART achieves the best performance on both Web Search Overall ($84.0\%$) and Memory Overall ($69.38\%$), surpassing Claude-Sonnet-4.6 by $+3.0\%$ and $+1.10$\% respectively. The memory advantage over GPT-5.4 is more substantial, with HEART outperforming by $+13.47$\% on Memory Overall, including gains of $+11.05\%$, $+10.39\%$, and $+18.97\%$ on KV, Vector, and Recursive Summarization. 
SFT-based models fail almost entirely: Hammer2.1-7B scores $0.0\%$ across all metrics, and the strongest SFT baseline xLAM-2-32B-FC-R reaches only $25.5$\%/$20.86$\%, showing no generalization to dynamic agentic tool scenarios.

\subsection{Ablation Study}

\begin{wraptable}{r}{0.5\linewidth}
\vspace{-12pt}
\centering
\caption{Ablation study on component contribution evaluated on ToolBench and NESTFUL.}
\vspace{-6pt}
\label{tab:ablation_component}
\resizebox{1\linewidth}{!}{%
\begin{tabular}{lcc|cc}
\toprule
\multirow{2}{*}{\textbf{Variant}} 
& \multicolumn{2}{c|}{\textbf{ToolBench}} 
& \multicolumn{2}{c}{\textbf{NESTFUL}} \\
\cmidrule(lr){2-3} \cmidrule(lr){4-5}
& \textbf{Avg. Pass} & \textbf{Win Rate} 
& \textbf{Full Acc.} & \textbf{Win Rate} \\
\midrule
w/o Planner   & 57.2 & 64.1 & 0.26 & 0.51 \\
w/o Router    & 62.4 & 67.3 & 0.28 & 0.52 \\
w/o ToolFace  & 16.1 & 27.9 & 0.06 & 0.06 \\
w/o Verifier  & 47.6 & 49.6 & 0.15 & 0.33 \\
\midrule
\textbf{HEART} 
& \textbf{75.1} & \textbf{75.7} 
& \textbf{0.44} & \textbf{0.75} \\
\bottomrule
\end{tabular}}
\vspace{-12pt}
\end{wraptable}

\textbf{Component Contribution}.
We evaluate four HEART ablations on ToolBench and NESTFUL, each removing one core component: (1) w/o Planner, which removes intent analysis and information sufficiency checking, passing the raw user query directly to the Router without structured decomposition or clarification;(2) w/o Router, which removes parameter mapping and tool dispatch, instead having the Planner invoke Tool Primitives directly from the invocation plan without intermediate argument resolution; (3) w/o ToolFace \& Tool Primitives, which removes the centralized schema--function registry and the natural language invocation abstraction, exposing raw API schemas directly to the Router and dispatching calls; and (4) w/o Verifier, which removes the verification loop, so execution results are returned to the user without evaluation or feedback-driven re-planning. Table~\ref{tab:ablation_component} reports results.

% \begin{table}[htbp]
% \centering
% \caption{Ablation study on component contribution evaluated on ToolBench and NESTFUL.}
% \label{tab:ablation_component}
% \resizebox{0.5\linewidth}{!}{%
% \begin{tabular}{lcc|cc}
% \toprule
% \multirow{2}{*}{\textbf{Variant}} 
% & \multicolumn{2}{c|}{\textbf{ToolBench}} 
% & \multicolumn{2}{c}{\textbf{NESTFUL}} \\
% \cmidrule(lr){2-3} \cmidrule(lr){4-5}
% & \textbf{Avg. Pass} & \textbf{Win Rate} 
% & \textbf{Full Acc.} & \textbf{Win Rate} \\
% \midrule
% w/o Planner                     & 57.2 & 64.1 & 0.26 & 0.51 \\
% w/o Router                      & 62.4 & 67.3 & 0.28 & 0.52 \\
% w/o ToolFace & 16.1 & 27.9 & 0.06 & 0.06 \\
% w/o Verifier                    & 47.6 & 49.6 & 0.15 & 0.33 \\
% \midrule
% \textbf{HEART}                  & \textbf{75.1} & \textbf{75.7} 
%                                 & \textbf{0.44} & \textbf{0.75} \\
% \bottomrule
% \end{tabular}}
% \end{table}

The most dramatic degradation is observed under \textbf{w/o ToolFace \& Tool Primitives}, which reduces ToolBench Avg. Pass Rate from $75.1$ to $16.1$ and NESTFUL Full Acc. from $0.44$ to $0.06$, confirming that the natural language 
interface for tools is the most critical component in HEART. Removing the Verifier produces the second-largest drop (Pass Rate $47.6$, Full Acc. $0.15$), demonstrating that feedback-driven re-planning is essential for recovering from execution failures. The w/o Planner and w/o Router variants show moderate but consistent degradation, confirming that intent decomposition and parameter mapping both contribute necessary intermediary layers to the pipeline.

\begin{wraptable}{r}{0.52\linewidth}
\centering
\vspace{-13pt}
\caption{Effect of re-planning budget $B$ on ToolBench and NESTFUL.}
\vspace{-6pt}
\label{tab:ablation_budget}
\resizebox{1\linewidth}{!}{%
\begin{tabular}{lcc|cc}
\toprule
\multirow{2}{*}{\textbf{Budget $B$}} 
& \multicolumn{2}{c|}{\textbf{ToolBench}} 
& \multicolumn{2}{c}{\textbf{NESTFUL}} \\
\cmidrule(lr){2-3} \cmidrule(lr){4-5}
& \textbf{Avg. Pass} & \textbf{Win Rate} 
& \textbf{Full Acc.} & \textbf{Win Rate} \\
\midrule
$B = 0$ & 47.6 & 49.6 & 0.15 & 0.33 \\
$B = 1$ & 51.4 & 57.9 & 0.27 & 0.49 \\
$B = 2$ & 65.9 & 66.8 & 0.42 & 0.68 \\
$B = 3$ & 75.1 & 75.7 & 0.44 & 0.75 \\
$B = 5$ & 75.3 & 76.2 & 0.44 & 0.75 \\
\bottomrule
\end{tabular}}
\vspace{-15pt}
\end{wraptable}

\textbf{Re-planning Budget.} We study the effect of the maximum re-planning budget $B$ by varying $B \in \{0, 1, 2, 3, 5\}$ and evaluating on ToolBench and NESTFUL, where $B = 0$ disables re-planning entirely, reducing HEART to a single-pass execution pipeline without feedback-driven recovery.

% \begin{table}[htbp]
% \centering
% \caption{Effect of re-planning budget $B$ on ToolBench and NESTFUL. 
% $B = 0$ disables re-planning entirely.}
% \label{tab:ablation_budget}
% \resizebox{0.5\linewidth}{!}{%
% \begin{tabular}{lcc|cc}
% \toprule
% \multirow{2}{*}{\textbf{Budget $B$}} 
% & \multicolumn{2}{c|}{\textbf{ToolBench}} 
% & \multicolumn{2}{c}{\textbf{NESTFUL}} \\
% \cmidrule(lr){2-3} \cmidrule(lr){4-5}
% & \textbf{Avg. Pass} & \textbf{Win Rate} 
% & \textbf{Full Acc.} & \textbf{Win Rate} \\
% \midrule
% $B = 0$ & 47.6 & 49.6 & 0.15 & 0.33 \\
% $B = 1$ & 51.4 & 57.9 & 0.27 & 0.49 \\
% $B = 2$ & 65.9 & 66.8 & 0.42 & 0.68 \\
% $B = 3$ & 75.1 & 75.7 & 0.44 & 0.75 \\
% $B = 5$ & 75.3 & 76.2 & 0.44 & 0.75 \\
% \bottomrule
% \end{tabular}}
% \end{table}

Table~\ref{tab:ablation_budget} reports results. Performance improves consistently from $B = 0$ to $B = 3$ across both benchmarks, with the largest gains concentrated between $B = 1$ and $B = 2$: ToolBench Pass Rate jumps from $51.4$ to $65.9$ and NESTFUL Full Acc. from $0.27$ to $0.42$, indicating that the second re-planning round resolves the majority of recoverable failures. 
% Increasing the budget to $B = 5$ yields negligible further improvement ($+0.2$ Pass Rate, $+0.5$ Win Rate on ToolBench), confirming that performance saturates at $B = 3$. We therefore adopt $B = 3$ as the default, which achieves the best trade-off between task completion and inference cost.
$B=5$ yields negligible further improvement, confirming saturation at $B=3$, which we adopt as the default.

\subsection{Case Study}

\begin{wraptable}{r}{0.55\linewidth}
\vspace{-12pt}
\centering
\caption{Task Completion Rate (\%) on 50 tasks across five domains. Each domain contains 10 tasks evaluated by human annotators.}
\vspace{-5pt}
\label{tab:realworld}
\resizebox{1\linewidth}{!}{%
\begin{tabular}{lcccc}
\toprule
\textbf{Domain} & \textbf{GPT-5.4} & \textbf{Claude-4.6-Sonnet} 
& \textbf{Gemini-3.1-Pro} & \textbf{HEART} \\
\midrule
Travel Planning  & 20 & 20 & 10 & \textbf{80} \\
Healthcare       & 20 & 30 & 30 & \textbf{70} \\
E-commerce       & 10 & 10 & 20 & \textbf{80} \\
Finance          & 30 & 30 & 30 & \textbf{100} \\
Local Services   & 20 & 30 & 20 & \textbf{90} \\
\midrule
\textbf{Overall} & 20 & 24 & 22 & \textbf{84} \\
\bottomrule
\end{tabular}}
\vspace{-12pt}
\end{wraptable}

\textbf{End-to-End Real World Task}.
To evaluate HEART in realistic deployment conditions, we construct a benchmark of 50 end-to-end real-world tasks spanning five domains: \textit{Travel Planning}, \textit{Healthcare}, \textit{E-commerce}, \textit{Finance}, and \textit{Local Services} with 10 tasks per domain. Each task is grounded in a concrete user persona (e.g., a student booking a flight, a group organizer finding catering) and expressed as a request requiring multi-step tool invocation to resolve. Tasks range from single-tool lookups (e.g., retrieving gas prices) to multi-tool workflows involving search, comparison, and conditional booking across heterogeneous APIs.

% \begin{table}[t]
% \centering
% \caption{Task Completion Rate (TCR) on the 50 End-to-End Real-World Tasks across five domains. Each domain contains 10 tasks evaluated by human annotators.}
% \label{tab:realworld}
% \resizebox{0.7\linewidth}{!}{%
% \begin{tabular}{lcccc}
% \toprule
% \textbf{Domain} & \textbf{GPT-5.4} & \textbf{Claude-4.6-Sonnet} 
% & \textbf{Gemini-3.1-Pro} & \textbf{HEART} \\
% \midrule
% Travel Planning  & 20 & 20 & 10\% & \textbf{80} \\
% Healthcare       & 20 & 30 & 30\% & \textbf{70} \\
% E-commerce       & 10 & 10 & 20\% & \textbf{80} \\
% Finance          & 30 & 30 & 30\% & \textbf{100} \\
% Local Services   & 20 & 30 & 20\% & \textbf{90} \\
% \midrule
% \textbf{Overall} & 20 & 24 & 22 & \textbf{84} \\
% \bottomrule
% \end{tabular}}
% \end{table}

% From Table~\ref{tab:realworld}, HEART achieves an overall TCR of $0.84$, outperforming GPT-5.4 ($0.20$) by $+64\%$, Claude-4.6-Sonnet ($0.24$) by $+60\%$, and Gemini-3.1-Pro ($0.22$) by $+62\%$. The consistently low TCR of commercial models reveals a critical gap between \textit{recommendation} and \textit{execution}: all three baselines reliably retrieve and suggest relevant options, but fail when tasks require acting on behalf of the user — populating form fields, binding reservation parameters, or completing multi-step transactional workflows. HEART supports grounded argument mapping and stateful execution, enabling it to complete tasks that require acting in the world, not describing it.

Table~\ref{tab:realworld} shows that HEART achieves $84$\% task completion, outperforming GPT-5.4 ($20$\%), Claude-4.6-Sonnet ($24$\%), and Gemini-3.1-Pro ($22$\%) by over $60$\%. The gap reveals a critical divide between recommendation and execution: commercial models reliably retrieve options but fail when tasks require acting on behalf of the user, whereas HEART's grounded argument mapping and stateful execution enable it to complete tasks that require acting in the world.

\begin{wraptable}{r}{0.55\linewidth}
\centering
\vspace{-12pt}
\caption{Tool selection accuracy (ACC\%) and attack success rate (ASR\%) on ToolBench. HEART achieves $0.0\%$ ASR under both attack strategies because tool schemas are stored in ToolFace and invoked via Tool Primitives, never appearing in the LLM's input prompt and thus eliminating the attack surface by construction.}
\label{tab:injection}
\vspace{-5pt}
\resizebox{\linewidth}{!}{%
\begin{tabular}{llcccc}
\toprule
\textbf{Attack} & \textbf{Metric} & \textbf{GPT-5.4} & \textbf{Claude-4.6-Sonnet} & \textbf{Gemini-3.1} & \textbf{HEART} \\
\midrule
No Attack       & ACC  & 99.0 & 99.0 & 98.6 & \textbf{--} \\
Gradient-Free   & ASR & 82.2 & 74.4 & 78.8 & \textbf{0.0} \\
Gradient-Based  & ASR  & 70.6 & 66.0 & 72.2 & \textbf{0.0} \\
\bottomrule
\end{tabular}}
\vspace{-15pt}
\end{wraptable}

\textbf{Robustness against Prompt Injection Attack.}
Existing retrieve-then-select paradigms expose tool metadata in the model's input prompt, creating an attack surface for prompt injection attacks. Following ToolHijacker~\cite{shi2025prompt}, we consider an attacker who injects a malicious tool document into the retrieved candidate set to manipulate the agent's tool selection, and report ACC under no attack and ASR under Gradient-Free and Gradient-Based strategies on ToolBench to evaluate Robustness under such attacks.
% \begin{table}[htbp]

As shown in Table~\ref{tab:injection}, all models remain highly vulnerable, with ASR reaching up to $70\%$ under Gradient-Free attack. HEART achieves $0.0\%$ ASR under both strategies. Tool schemas are stored in ToolFace and executed via Tool Primitives, never appearing in the LLM's input prompt. The attack surface that ToolHijacker exploits does not exist.
% \subsection{Discussion}
% \label{sec:discussion}

% The results across five benchmarks empirically validate the three design properties of Tool Primitives introduced in Section~\ref{Methodology}. \textbf{(1) Natural language invocation.} Each Primitive accepts a natural language request and resolves the schema internally, freeing the Router from raw parameter formatting. Removing this abstraction drops ToolBench Pass Rate from $75.1$ to $16.1$ (Table~\ref{tab:ablation_component}), the largest degradation among all components. As a side effect, schemas never enter the LLM's context, yielding $0.00\%$ ASR under prompt injection (Table~\ref{tab:injection}). \textbf{(2) Inter-tool communication.} The result of one Primitive is passed as natural language context to the next, enabling LLM-to-LLM dialogue across dependent steps. On NESTFUL (Table~\ref{NESTEDFUL_compare}), SFT models collapse to $0.00$ Full Acc. and DeepSeek-V3 (685B) reaches only $0.09$, while HEART attains $0.44$---neither fine-tuning nor scale substitutes for this interface. \textbf{(3) Execution isolation.} Each Primitive validates arguments internally and surfaces failures as structured signals to the Verifier, localizing errors rather than propagating them. This makes feedback-driven recovery tractable: HEART sustains Pass$^4$ on $\tau^2$-Bench (Table~\ref{t2_compare}) where commercial baselines degrade, and most recoverable failures resolve within $B=2$ rounds (Table~\ref{tab:ablation_budget}).

% \vspace{-7pt}
\subsection{Discussion}
% \vspace{-7pt}
\label{sec:discussion}
The results validate the three design properties of Tool Primitives. \textbf{(1) Natural language invocation.} Each Primitive resolves the schema internally, freeing the Router from raw parameter formatting. Removing this drops ToolBench Pass Rate from $75.1$ to $16.1$ (Table~\ref{tab:ablation_component}), and yields $0.00\%$ ASR under prompt injection (Table~\ref{tab:injection}). \textbf{(2) Inter-tool communication.} Primitive results are passed as context to the next, enabling LLM-to-LLM dialogue across dependent steps. On NESTFUL (Table~\ref{NESTEDFUL_compare}), SFT models collapse to $0.00$ Full Acc. and DeepSeek-V3 reaches only $0.09$, while HEART attains $0.44$. \textbf{(3) Execution isolation.} Each Primitive surfaces failures as signals to the Verifier, localizing errors. HEART sustains Pass$^4$ on $\tau^2$-Bench (Table~\ref{t2_compare}) where baselines degrade, and most failures resolve within $B=3$ rounds (Table~\ref{tab:ablation_budget}). We also discuss the limitations of HEART in Appendix~\ref{limitation}.

% \section{Discussion}\label{disscusion}
% \input{Secs/discussion}

% \vspace{-7pt}
\section{Conclusion}\label{con}
% \vspace{-7pt}
We propose HEART, a harness engineering framework that addresses key limitations of existing LLM tool-use systems, including scalability to large tool catalogs, robustness in multi-turn interaction, compositional reasoning, and failure recovery. By introducing Tool Primitives and a coordinated multi-agent design with planning, routing, and verification, HEART transforms tool use into a structured and iterative reasoning process. Experiments across multiple benchmarks show that HEART consistently improves performance, efficiency, and robustness over existing approaches, including strong resilience to real-world challenges such as prompt injection. We hope our proposed HEART framework can be the real heart of the LLM tool use.

% \section*{References}

% \newpage
\bibliographystyle{unsrt}
\bibliography{ref}

%%%%%%%%%%%%%%%%%%%%%%%%%%%%%%%%%%%%%%%%%%%%%%%%%%%%%%%%%%%%
\newpage
\appendix
\section{Detailed Agent Role Specifications}
\label{appendix:agent-roles}

We expand on the role specifications of the three collaborative agents introduced in the main text.

\subsection{Planner}
The Planner serves as HEART's entry point, responsible for two sequential functions: \textbf{intent analysis} and \textbf{information sufficiency checking}. Upon receiving a user query $q$, the Planner first decomposes $q$ into a structured intent representation that identifies the target task, relevant tool categories, and any explicit constraints or preferences expressed by the user. It then evaluates whether the current context $\mathcal{C}_t$—comprising the user query, any previously acquired information, and execution history—provides sufficient information to construct a fully specified invocation plan.

If the sufficiency check returns \texttt{insufficient}, the Planner generates a targeted clarification request $c_t$ and returns it to the user, initiating another interaction round. This loop continues until the context is judged sufficient, at which point the Planner produces an invocation plan $\Pi = (\pi_1, \pi_2, \ldots, \pi_K)$, an ordered sequence of $K$ tool invocation steps, each $\pi_k$ specifying the tool to be invoked at step $k$ to address the user's request.

\subsection{Router}
Given an invocation plan $\Pi$ from the Planner, the Router is responsible for \textbf{parameter mapping} and \textbf{tool dispatch}. For each step $\pi_k \in \Pi$, which specifies the target Tool Primitive but contains no concrete argument values, the Router resolves the required arguments from the current context $\mathcal{C}_t$ and constructs a natural language invocation request $x_k$ that encodes the target tool, the intended operation, and the resolved parameter values, along with execution hyperparameters \texttt{H-Params}$_k$ such as retry count and priority level.

Note that $\mathcal{C}_t$ serves a different role here than in the Planner: whereas the Planner consults $\mathcal{C}_t$ to assess information sufficiency, the Router uses $\mathcal{C}_t$ purely as a value source for parameter resolution, ensuring the two components are complementary.

Directly dispatching from the Router to raw tool implementations faces two persistent limitations. First, precise parameter formatting is fragile: even when the Router correctly identifies the target tool and user intent, small deviations in argument type, naming, or value encoding can cause schema validation failures. Second, with large tool catalogues, the Router must reason over raw API schemas—a verbose, low-level representation that degrades tool selection accuracy and increases the likelihood of incomplete or malformed bindings. These limitations motivate the Tool Primitive abstraction, which absorbs schema resolution and argument binding within each Primitive. The Router thus dispatches $x_k$ to the corresponding Tool Primitive $\mathcal{P}_k$, which internally validates and maps the described arguments against schema $s_k$, and executes the underlying function $f_k$.

\subsection{Verifier}
The Verifier evaluates each execution result $r_k$ returned by a Tool Primitive against four structured criteria before the result is accepted or escalated for re-planning. Detailed criterion definitions are provided in Appendix~\ref{appendix:verifier-criteria}.

If verification passes, execution proceeds to the next step $\pi_{k+1}$ or, if $k = K$, the final result is returned to the user. If verification fails, the Verifier generates structured feedback $\phi_k$ that diagnoses which criterion failed and why, and returns $\phi_k$ to the Planner to trigger targeted re-planning with the enriched context.

\section{Verifier Evaluation Criteria}
\label{appendix:verifier-criteria}

The Verifier evaluates each execution result $r_k$ along four structured criteria:

\begin{itemize}
    \item \textbf{Task Completion}: Does $r_k$ satisfy the objective specified in invocation step $\pi_k$?
    
    \item \textbf{Argument Consistency}: Are the arguments resolved by the Router consistent with the user's original intent and the constraints encoded in schema $s_k$?
    
    \item \textbf{Execution Validity}: Did the tool execute without runtime errors, and does the return value conform to the expected output schema?
    
    \item \textbf{Constraint Satisfaction}: Are any task-level or domain-specific constraints—such as rate limits, access permissions, or business rules—respected by the execution outcome?
\end{itemize}

Each criterion is assessed independently; failure in any single criterion yields $v_k = \texttt{fail}$ and triggers feedback generation. This decomposition enables the Verifier to produce actionable, criterion-specific diagnostics rather than opaque pass/fail signals.

\section{End-to-End Execution Flow}\label{end-to-end}

Algorithm~\ref{alg:heart} summarizes the full HEART pipeline. Starting from user query $q$, the Planner iteratively acquires missing information until context is sufficient, then produces invocation plan $\Pi$. The Router sequentially binds and dispatches each step, passing inter-step results through Tool Primitive interfaces. The Verifier evaluates each result and either advances execution or returns structured feedback for re-planning. This loop continues until all $K$ steps pass verification or a maximum re-planning budget is exhausted.

\begin{algorithm}[htbp]
\caption{Pseudo-code of HEART}
\label{alg:heart}
\begin{algorithmic}[1]
\Require User query $q$, ToolFace $\mathcal{T}$, max re-plan budget $B$
\Ensure Final execution result or structured failure report
\State $\mathcal{C} \leftarrow \{q\}$, $b \leftarrow 0$
\While{$\textsc{Planner}(q, \mathcal{C}) = \texttt{insufficient}$}
    \State $c \leftarrow \textsc{Planner}.\textsc{Clarify}(q, \mathcal{C})$
    \State $\mathcal{C} \leftarrow \mathcal{C} \cup \{\textsc{User}(c)\}$
\EndWhile
\State $\Pi \leftarrow \textsc{Planner}.\textsc{Plan}(q, \mathcal{C})$
\While{$b \leq B$}
    \For{each step $\pi_k \in \Pi$}
        \State $\beta_k \leftarrow \textsc{Router}(\pi_k, \mathcal{C})$
        \State $r_k \leftarrow \mathcal{P}_k(\beta_k \mid r_{k-1})$
        \If{$\textsc{Verifier}(r_k, \pi_k, s_k) = \texttt{fail}$}
            \State $\phi_k \leftarrow \textsc{Verifier}.\textsc{Feedback}(r_k, \pi_k, s_k)$
            \State $\mathcal{C} \leftarrow \mathcal{C} \cup \{\phi_k\}$
            \State $\Pi \leftarrow \textsc{Planner}.\textsc{Plan}(q, \mathcal{C})$
            \State $b \leftarrow b + 1$; \textbf{break}
        \EndIf
    \EndFor
    \State \Return $r_K$
\EndWhile
\State \Return \textsc{FailureReport}$(\mathcal{C})$
\end{algorithmic}
\end{algorithm}

\section{Prompt Templates}\label{prompt}
\subsection{Planner Prompt}

\begin{tcolorbox}[title=Planner System Prompt, colback=purple!5, colframe=purple!60, fonttitle=\bfseries, breakable]
\textbf{Role.} You are the Planner in the tool-calling framework. Your responsibilities are (1) intent analysis and (2) information sufficiency checking.

\textbf{Intent Analysis.} Given the user query, identify:
\begin{itemize}
  \item The target task and its objective.
  \item The relevant tool categories required to fulfill the task.
  \item Any explicit constraints or preferences expressed by the user (e.g., priority, domain, format).
\end{itemize}

\textbf{Information Sufficiency Checking.} Determine whether the current context provides all information needed to construct a complete invocation plan. Output one of:
\begin{itemize}
  \item \texttt{SUFFICIENT} — proceed to generate an invocation plan.
  \item \texttt{INSUFFICIENT} — generate a targeted clarification question asking only for the missing information.
\end{itemize}

\textbf{Invocation Plan Format.} When context is sufficient, output an ordered JSON list of invocation steps:
\begin{verbatim}
{
  "status": "SUFFICIENT",
  "plan": [
    {
      "step": 1,
      "tool_category": "<category>",
      "tool_hint": "<tool name or description>",
      "objective": "<what this step should accomplish>",
      "dependencies": []
    },
    {
      "step": 2,
      ...
      "dependencies": [1]
    }
  ]
}
\end{verbatim}

\textbf{Clarification Format.} When context is insufficient:
\begin{verbatim}
{
  "status": "INSUFFICIENT",
  "clarification": "<single targeted question>"
}
\end{verbatim}

\textbf{Re-planning.} When provided with Verifier feedback, revise the plan to address the diagnosed failure. Output a revised plan in the same format, annotated with \texttt{``replanned'': true}.
\end{tcolorbox}
\subsection{Router Prompt}
\begin{tcolorbox}[title=Router System Prompt, colback=blue!5, colframe=blue!60, fonttitle=\bfseries, breakable]
\textbf{Role.} You are the Router in the tool-calling framework. Your responsibilities are (1) parameter mapping and (2) tool dispatch. 

\textbf{Parameter Mapping.} For each step in the invocation plan, given the current context (the user query, prior clarifications, and results of previously executed steps):
\begin{itemize}
  \item Resolve the value of every required parameter of the step from the context. Values may originate from the original user query, earlier clarification turns, or the structured result of an upstream step (referenced as \texttt{\$step\_j.<field>}).
  \item Group all resolved values that belong to the same step together; do not mix values across steps.
  \item Encode the resolved bindings into a single natural-language invocation request that names the target Tool Primitive, states the intended operation, and embeds the resolved values inline. Do not emit a raw schema-compliant parameter dictionary --- the Tool Primitive performs schema-level argument binding internally.
  \item Never invent a value that is not grounded in the context, and never override the target Tool Primitive chosen by the Planner.
\end{itemize}

\textbf{Tool Dispatch.} Attach execution-level hyperparameters for each step:
\begin{itemize}
  \item \texttt{retry}: integer in $[1, 5]$. Use higher values for safety- or finance-critical actions and lower values for read-only lookups.
  \item \texttt{priority}: one of \{\texttt{low}, \texttt{normal}, \texttt{high}\}. Use \texttt{high} when the user signals urgency or when the step lies on the critical path of a time-sensitive task.
  \item \texttt{timeout\_s}: integer seconds; default 30.
\end{itemize}

\textbf{Dispatch Format.} Output an ordered JSON list, one entry per step in the invocation plan, in the original step order:
\begin{verbatim}
{
  "status": "READY",
  "dispatch": [
    {
      "step": 1,
      "target_tool": "<tool name or tool hint>",
      "invocation_request": "<natural-language request>",
      "resolved_bindings": { "<param>": "<value>", ... },
      "h_params": {
        "retry": <int>,
        "priority": "<low|normal|high>",
        "timeout_s": <int>
      }
    },
    {
      "step": 2,
      ...
    }
  ]
}
\end{verbatim}
\textbf{Cross-Step Consistency.} When two or more steps share the same parameter (e.g., the same card identifier or user ID), resolve the value once from the context and reuse it consistently across all affected steps' bindings.
\end{tcolorbox}

\subsection{Tool Primitive Prompt}
\begin{tcolorbox}[title=Tool Primitive System Prompt, colback=green!5, colframe=green!60, fonttitle=\bfseries, breakable]
\textbf{Role.} You are a Tool Primitive in the tool-calling framework. You wrap a single tool from ToolFace and serve as its agent-native interface. You are instantiated with two pieces of information: the tool's \textbf{schema} (specifying parameter names, types, constraints, and return format) and the tool's executable \textbf{function}. 

\textbf{Schema Resolution.} Given the natural-language invocation request from the Router and any optional context (e.g., the structured result of a prior Tool Primitive passed as upstream context):
\begin{itemize}
  \item Interpret the request to extract the intended operation and the values described for each parameter.
  \item Map each described value to the argument space defined by your schema, performing type coercion (e.g., string-to-integer, date normalization), enum matching, and unit conversion as needed.
  \item Apply schema-level constraints: required-field checks, value-range validation, format validation (e.g., regex patterns), and any default values specified by the schema.
  \item If the request references upstream context, extract the relevant field from that context and bind it to the corresponding parameter.
\end{itemize}

\textbf{Function Execution.} Once all arguments are resolved and validated, invoke the underlying function with the bound argument dictionary. Capture the raw return value, runtime errors (if any), and execution metadata (e.g., latency, status code).

\textbf{Success Format.} If schema resolution succeeds and the function returns without runtime error, output:
\begin{verbatim}
{
  "status": "SUCCESS",
  "tool": "<tool name from schema>",
  "bound_arguments": { "<param>": "<resolved value>", ... },
  "result": { ... },
  "summary": "<one-sentence natural-language summary of the result>"
}
\end{verbatim}
The \texttt{summary} field provides a natural-language rendering of the result so that it can be passed as upstream context to a downstream Tool Primitive without requiring the Router or Planner to manage intermediate state.

\textbf{Failure Format.} If schema resolution fails (e.g., a required parameter cannot be extracted from the request, a value violates a schema constraint, or type coercion is impossible), or if the function raises a runtime error, do not retry silently. Surface a structured failure to the Verifier:
\begin{verbatim}
{
  "status": "FAILURE",
  "tool": "<tool name from schema>",
  "failure_type": "<SCHEMA_RESOLUTION
                    | CONSTRAINT_VIOLATION
                    | RUNTIME_ERROR>",
  "details": "<which parameter or constraint failed, and why>",
  "bound_arguments": { "<param>": "<resolved value>", ... }
}
\end{verbatim}

\textbf{Execution Isolation.} Errors must remain local to this Tool Primitive --- never call another tool, never modify the invocation plan, and never attempt to recover from a failure on your own. The Verifier will diagnose the failure and trigger re-planning if needed.
\end{tcolorbox}

\subsection{Verifier Prompt}
\begin{tcolorbox}[title=Verifier System Prompt, colback=orange!10, colframe=orange!70, fonttitle=\bfseries, breakable]
\textbf{Role.} \textbf{Role.} You are the Verifier in the tool-calling framework. Your responsibilities are (1) evaluation of the execution result against four criteria and (2) feedback generation for the Planner when evaluation fails. You receive the execution result, the corresponding step in the invocation plan, and the tool's schema.

\textbf{Evaluation Criteria.} Assess the execution result against four criteria:
\begin{enumerate}
  \item \textbf{Task Completion}: Does the result satisfy the objective specified in the step? Is the output semantically aligned with what the user requested?
  \item \textbf{Argument Consistency}: Are the resolved arguments consistent with the user's original intent and the constraints encoded in the tool's schema? Were any required fields omitted or incorrectly substituted?
  \item \textbf{Execution Validity}: Did the tool execute without runtime errors? Does the return value conform to the expected output schema?
  \item \textbf{Constraint Satisfaction}: Are task-level or domain-specific constraints respected (e.g., rate limits, access permissions, business rules)?
\end{enumerate}

\textbf{Pass Format.} Issue \texttt{PASS} only when all four criteria are satisfied:
\begin{verbatim}
{
  "status": "PASS",
  "step": <int>,
  "criteria": {
    "task_completion":         "pass",
    "argument_consistency":    "pass",
    "execution_validity":      "pass",
    "constraint_satisfaction": "pass"
  }
}
\end{verbatim}

\textbf{Fail Format.} If any criterion fails, return structured feedback to the Planner:
\begin{verbatim}
{
  "status": "FAIL",
  "step": <int>,
  "criteria": {
    "task_completion":         "pass" | "fail",
    "argument_consistency":    "pass" | "fail",
    "execution_validity":      "pass" | "fail",
    "constraint_satisfaction": "pass" | "fail"
  },
  "feedback": "<structured diagnosis identifying which 
               criterion failed, why it failed, and what 
               information the Planner needs to re-plan 
               successfully>"
}
\end{verbatim}

\textbf{Feedback Guidelines.}
\begin{itemize}
  \item Feedback must be actionable: specify the failure mode precisely (e.g., ``wrong argument type for \texttt{card\_id}: expected \texttt{int}, received \texttt{str}'') rather than vague descriptions.
  \item Do not speculate about failures not evidenced in the execution result or the schema.
  \item When multiple criteria fail simultaneously, list all failed criteria and prioritize the root cause in the diagnosis.
\end{itemize}
\end{tcolorbox}

\section{Additional Experiments}

\subsection{Model Pricing}
\label{appendix:pricing}

Table~\ref{tab:pricing} lists the input/output token prices for all models used in our cost analysis.

\begin{table}[htbp]
\centering
\caption{Token pricing for models used in $\tau^2$-Bench cost comparison (USD per 1M tokens).}
\label{tab:pricing}
\resizebox{\linewidth}{!}{%
\begin{tabular}{lccc}
\toprule
\textbf{Model} & \textbf{Input (\$/1M)} & \textbf{Output (\$/1M)} & \textbf{Source} \\
\midrule
Qwen3-8B          & 0.18  & 0.70  & \url{https://artificialanalysis.ai/models/qwen3-8b-instruct/} \\
GPT-5.4           & 2.50  & 15.00 & \url{https://developers.openai.com/api/docs/pricing} \\
GPT-4.1           & 2.00  & 8.00  & \url{https://developers.openai.com/api/docs/pricing} \\
GPT-4.1-mini      & 0.40  & 1.60  & \url{https://developers.openai.com/api/docs/pricing} \\
GPT-o4-mini       & 0.15  & 0.60  & \url{https://developers.openai.com/api/docs/pricing} \\
Claude-4.6-Sonnet & 3.00  & 15.00 & \url{https://claude.com/pricing\#api} \\
Gemini-3.1-Pro        & 2.00  & 12.00 & \url{https://ai.google.dev/gemini-api/docs/pricing\#gemini-3.1-pro-preview} \\
Llama-3.1-8B      & 0.02    & 0.05  & \url{https://novita.ai/models/model-detail/meta-llama-llama-3.1-8b-instruct} \\
\bottomrule
\end{tabular}}
\end{table}
\subsection{Additional ACEBench Comparisons}
\label{appendix:acebench-open}

We additionally compare HEART against open-source models and SFT-based tool-calling baselines on ACEBench in Table~\ref{ACEBench_compare_2}. HEART's Overall score of $86.9$\% surpasses the best open-source baseline Qwen2.5-Coder-32B-Instruct ($79.6$\%) by $+7.3$\%. The gap is most pronounced on Multi-Turn ($83.5$\% vs. $71.0$\%, $+12.5$\%) and Agent ($72.3$\% vs. $60.8$\%, $+11.5$\%). SFT-based tool-calling models struggle severely on Special and Agent categories: Watt-Tool-8B reaches only $6.0$\% on Special and $2.8$\% on Agent, and xLAM-7B-r scores $0.0$ on Preference, confirming that fine-tuning on narrow tool-use trajectories does not generalize to robustness-critical or agentic scenarios.

\begin{table*}[htbp]
\centering
\caption{Performance comparison (\%) of different methods across domains on the ACEBench.}

\Huge
\label{ACEBench_compare_2}
\resizebox{\linewidth}{!}{%
\begin{tabular}{lcccccc|ccc}
\toprule
\multirow{2}{*}{\textbf{Model}} 
& \multicolumn{6}{c}{\textbf{Normal}} 
& \multirow{2}{*}{\textbf{Special}} 
& \multirow{2}{*}{\textbf{Agent}} 
& \multirow{2}{*}{\textbf{Overall}} \\
\cmidrule(lr){2-7}
& \textbf{Atom} & \textbf{Single-Turn} & \textbf{Multi-Turn} & \textbf{Similar API} & \textbf{Preference} & \textbf{Summary} & & & \\
\midrule
Qwen2.5-Coder-32B-Instruct      & 90.2 & 81.0 & 71.0 & 83.0 & 81.0 & 84.1 & 80.7 & 60.8 & 79.6 \\
DeepSeek-V3                     & 91.5 & 84.0 & 77.0 & 83.0 & 83.0 & 86.5 & 73.0 & 34.5 & 74.8 \\
Qwen2.5-72B-Instruct            & 86.8 & 80.3 & 69.5 & 83.0 & 81.0 & 82.1 & 75.7 & 45.0 & 74.7 \\
Llama-3.1-70B-Instruct          & 82.5 & 68.3 & 63.5 & 79.0 & 68.0 & 75.5 & 38.3 & 42.3 & 60.4 \\
Qwen2.5-7B-Instruct             & 76.0 & 60.3 & 58.5 & 72.0 & 67.0 & 69.4 & 47.0 & 13.8 & 54.8 \\
DeepSeek-Coder-V2-Lite-Instruct & 75.2 & 57.8 & 46.5 & 72.0 & 65.0 & 66.4 & 40.3 &  2.0 & 49.5 \\
Qwen2.5-Coder-7B-Instruct       & 76.0 & 63.8 & 57.5 & 74.0 & 68.0 & 70.1 & 22.3 & 15.5 & 48.9 \\
Watt-Tool-8B                    & 85.7 & 69.3 & 55.5 & 79.0 & 64.0 & 75.6 &  6.0 &  2.8 & 45.7 \\
Hammer2.1-7B                    & 73.7 & 57.5 & 40.0 & 62.0 & 55.0 & 62.8 & 14.7 & 16.8 & 42.9 \\
Llama-3.1-8B-Instruct           & 51.9 & 39.8 & 28.0 & 66.0 & 46.0 & 46.6 & 21.0 &  5.3 & 33.4 \\
Phi-3-Mini-128k-Instruct        & 57.2 & 39.3 & 23.0 & 58.0 & 32.0 & 46.5 & 18.7 &  0.8 & 32.0 \\
xLAM-7B-r                       & 43.5 & 22.0 & 19.0 & 61.0 &  0.0 & 33.7 &  2.7 &  8.8 & 21.6 \\
Llama-3.2-3B-Instruct           & 38.7 & 15.3 &  9.0 & 42.0 & 32.0 & 29.6 &  9.4 &  0.0 & 19.6 \\
Hammer2.1-3B                    & 22.4 & 11.5 &  3.5 & 40.0 & 20.0 & 18.7 &  1.0 &  1.5 & 11.3 \\
\midrule
\textbf{HEART}                  & \textbf{94.2} & \textbf{87.5} & \textbf{83.5} & \textbf{89.0} & \textbf{86.0} & \textbf{89.8} & \textbf{95.0} & \textbf{72.3} & \textbf{86.9} \\
\bottomrule
\end{tabular}
}

\end{table*}

% \section{More Ablation Studies}
\subsection{Ablation Studies on Backbone LLM Scale}
Since HEART uses Qwen3-8B for all four roles (Planner, Router, Verifier, and Tool Primitives), we examine performance sensitivity to backbone capacity by evaluating four configurations on $\tau^2$-Bench: HEART (LLaMA-3.1-8B), which replaces Qwen3-8B with LLaMA-3.1-8B across all roles; HEART (Qwen3-8B), the default configuration; HEART (GPT-5.4), which uses GPT-5.4 across all roles; and HEART (GPT-5.4 + Qwen3-8B), which assigns GPT-5.4 to the Planner, Router, and Verifier while retaining Qwen3-8B for Tool Primitives.

\begin{table}[htbp]
\centering
\footnotesize
\caption{Ablation study on backbone LLM scale evaluated on $\tau^2$-Bench 
across three domains.}
\label{tab:ablation_scale}
\resizebox{\linewidth}{!}{%
\begin{tabular}{llcccc|cc}
\toprule
\textbf{Variant} & \textbf{Domain} 
& \textbf{Pass$^1$} & \textbf{Pass$^2$} & \textbf{Pass$^3$} & \textbf{Pass$^4$}
& \textbf{Tokens / Cost (\$)} & \textbf{Latency (s)} \\
\midrule
\multirow{3}{*}{HEART (LLaMA-3.1-8B)}
& Retail  & 0.58 & 0.54 & 0.52 & 0.50 & 23,247 / 0.00116 & 19.52 \\
& Airline & 0.52 & 0.48 & 0.45 & 0.42 & 25,812 / 0.00129 & 21.14 \\
& Telecom & 0.45 & 0.41 & 0.38 & 0.36 & 22,439 / 0.00112 & 26.43 \\

\midrule
\multirow{3}{*}{HEART (GPT-5.4)}
& Retail  & 0.86 & 0.82 & 0.80 & 0.79 & 20,456 / 0.30684 & 26.34 \\
& Airline & 0.78 & 0.74 & 0.72 & 0.70 & 22,891 / 0.34337 & 29.58 \\
& Telecom & 0.68 & 0.64 & 0.62 & 0.61 & 19,573 / 0.29359 & 38.78 \\
\midrule
\multirow{3}{*}{\shortstack[l]{HEART \\ (GPT-5.4 + Qwen3-8B)}}
& Retail  & 0.83 & 0.79 & 0.77 & 0.74 & 18,362 / 0.19666 & 23.46 \\
& Airline & 0.74 & 0.70 & 0.68 & 0.65 & 19,147 / 0.20506 & 26.18 \\
& Telecom & 0.63 & 0.61 & 0.66 & 0.63 & 17,824 / 0.19089 & 42.84 \\
\midrule
\multirow{3}{*}{HEART (Qwen3-8B)}
& Retail  & 0.82 & 0.79 & 0.77 & 0.73 & 21,684 / 0.0152 & 18.72 \\
& Airline & 0.72 & 0.69 & 0.65 & 0.63 & 24,531 / 0.0172 & 21.34 \\
& Telecom & 0.62 & 0.59 & 0.55 & 0.50 & 20,887 / 0.0146 & 37.95 \\
\bottomrule
\end{tabular}}

\end{table}

Table~\ref{tab:ablation_scale} reports results. HEART (GPT-5.4) achieves the highest Pass$^k$ scores across all domains, confirming that stronger backbone capacity benefits planning and routing. However, HEART (Qwen3-8B) achieves competitive performance at a fraction of the cost: it matches HEART (GPT-5.4 + Qwen3-8B) on most metrics while reducing cost by over $12\times$ relative to HEART (GPT-5.4). The HEART (GPT-5.4 + Qwen3-8B) configuration further demonstrates that Tool Primitives are modular and composable: by retaining Qwen3-8B solely for Tool Primitives while delegating planning, routing, and verification to GPT-5.4, the hybrid variant achieves performance close to the full GPT-5.4 configuration, confirming that Tool Primitives can be integrated with arbitrary backbone LLMs without architectural modification. This plug-and-play compatibility suggests that practitioners can independently 
scale the orchestration layer and the execution layer according to their cost and performance requirements. HEART (LLaMA-3.1-8B) lags behind HEART (Qwen3-8B) by a substantial margin across all domains and all $k$, indicating that model quality rather than parameter count is the primary driver of performance within the same size class. Taken together, these results confirm that Qwen3-8B strikes the best balance between task completion and inference cost, justifying its selection as the default backbone.

% Table~\ref{tab:ablation_scale} shows that stronger backbones improve performance, with HEART (GPT-5.4) achieving the highest Pass$^k$ scores. 
% Yet the default HEART (Qwen3-8B) remains competitive at a fraction of the cost, matching the hybrid configuration while reducing cost by over $12\times$. The HEART (GPT-5.4 + Qwen3-8B) configuration further demonstrates that Tool Primitives are modular and composable: by retaining Qwen3-8B solely for Tool Primitives while delegating planning, routing, and verification to GPT-5.4, the hybrid variant achieves performance close to the full GPT-5.4 configuration, confirming that Tool Primitives can be integrated with arbitrary backbone LLMs without architectural modification. This plug-and-play compatibility suggests that practitioners can independently scale the orchestration layer and the execution layer according to their cost and performance requirements.

\section{Use of AI Assistants}
\label{app:llm}

AI assistants were used as auxiliary tools for manuscript preparation, including language polishing, clarity improvement, organization, and limited experimental workflows. 
All experimental design, methodological decisions, analyses, reported results, and final content were reviewed and verified by the authors.

\section{Limitation}\label{limitation}
While HEART demonstrates strong performance across diverse benchmarks, we acknowledge several limitations. First, the multi-agent design introduces higher token consumption than single-model baselines, though the use of a lightweight 8B backbone keeps overall API cost low; practitioners with stricter latency budgets may need to consolidate roles. Second, ToolFace currently relies on manually authored schemas to ensure interface quality, and extending the registry to new domains requires additional curation effort; automating schema generation is a natural direction for future work. Third, our robustness evaluation focuses on prompt injection at the tool-selection stage following ToolHijacker; broader threat models, such as compromised tool outputs, remain to be studied. We view these as opportunities for further refinement rather than fundamental obstacles to the framework.

% \section{Code and Result}\label{Dataset}
% We will publish ToolFace, and the comprehensive results of Tool Primitives on the web. For detailed information, please visit the following link: ~\url{https://anonymous.4open.science/r/32D8}.

%%%%%%%%%%%%%%%%%%%%%%%%%%%%%%%%%%%%%%%%%%%%%%%%%%%%%%%%%%%%

% \newpage
% \input{checklist.tex}

\end{document}